\documentclass[%
nofootinbib,
 amsmath,amssymb,
 aps,
prm,
]{revtex4-2}

\usepackage{newtxtext,newtxmath}  
\usepackage{graphicx}
\usepackage{booktabs}
\usepackage{dcolumn} 
\usepackage{bm}
\usepackage{microtype}
\usepackage{xcolor}
\definecolor{darkblue}{rgb}{0,0,0.6}
\definecolor{apsblue}{rgb}{0.18,0.19,0.57}
\usepackage[colorlinks,linkcolor=apsblue,citecolor=apsblue,urlcolor=apsblue]{hyperref}

\newcommand{\MO}[1]{\textcolor{red}{#1}}
\newcommand{\DC}[1]{\textcolor{blue}{#1}}
\newcommand{\CL}[1]{\textcolor{purple}{#1}}

\newcommand{\R}[2]{R[{#1}, {#2}]}
\newcommand{\Rb}[2]{{\bm R}[{#1}, {#2}]}
\newcommand{\mean}[1]{{\mathbb{E}}[#1]}

\newcommand{\NS}{N_\mathcal{S}}

\graphicspath{{figures/}:{./}}

\begin{document}

\title{
The limits of interpretability in multiple linear regression
}


\author{Anand Sharma}
\affiliation{Indian Institute of Science Education and Research,
Dr. Homi Bhabha Road, Pashan, Pune 411008, India}

\author{Chen Liu}
\affiliation{Innovation and Research Division, Ge-Room Inc., 93160 Noisy le Grand, France}

\author{Daniele Coslovich}
\affiliation{Dipartimento di Fisica, Universit\`a di Trieste, Strada Costiera 11, 34151, Trieste, Italy}
\email{dcoslovich@units.it}

\author{Misaki Ozawa}
\affiliation{Univ. Grenoble Alpes, CNRS, LIPhy, 38000 Grenoble, France}
\email{misaki.ozawa@univ-grenoble-alpes.fr}

\begin{abstract}

Interpreting machine-learning models has attracted increasing attention, particularly in the physical sciences, where one often seeks to understand the underlying mechanisms rather than merely make predictions.
Multiple linear regression is often regarded as an interpretable alternative to more complex models, such as deep neural networks, because its predictions are expressed as explicit weighted sums of input features.
However, when input features are strongly correlated, namely in the presence of multicollinearity, the learned weights can exhibit large dataset-to-dataset fluctuations and oscillatory behavior across physically similar features, making their interpretation difficult or even impossible.
Although the instability of the weights under multicollinearity is well known in statistics, its consequences for physical interpretation, in particular its connection to oscillatory weights across physically similar features, have not been systematically clarified.
Here, we theoretically discuss the mechanism behind this loss of interpretability by analyzing the eigenmodes of the feature correlation matrix.
We show that small-eigenvalue modes associated with multicollinearity amplify fluctuations in the weights and generate oscillatory patterns that do not necessarily reflect meaningful contributions.
We test this theoretical picture numerically on physics datasets and show that Ridge regularization suppresses these unstable modes, although the resulting weights must still be interpreted with caution. We further confirm the generality of our findings beyond physics by analyzing a diverse collection of publicly available datasets. Our results clarify why, in the presence of multicollinearity, physical interpretation can remain difficult even for linear regression models. 
\end{abstract}

\maketitle


\section{Introduction}

While machine learning (ML) has achieved remarkable success in predicting complex phenomena
in several branches of physics~\cite{carleo2019machine}, it often fails to provide genuine physical understanding.
This is because many high-performing ML models, such as deep neural networks, are very complex and still difficult to interpret.
This limitation is now widely recognized across ML-related fields.
As emphasized by statements like ``prediction is not understanding''~\cite{teney2022predicting} or ``to predict is not to explain''~\cite{Thom_2016}, accurate predictions alone do not automatically lead to insight or intuition into the mechanisms underlying the phenomenon of interest. 
This issue is particularly relevant in the physical sciences,
where interpretability is not just a secondary requirement: it is essential for extracting physical principles, identifying relevant variables, and building simplified theoretical descriptions~\cite{brunton2016discovering,rudy2017data,Wetzel_Ha_Iten_Klopotek_Liu_2025,Rowan_Doostan_2025}.

It is often believed that there is a trade-off between predictive performance and interpretability~\cite{gunning2019xai} (see also Fig. 12 of Ref.~\cite{arrieta2020explainable}).
Deep learning models can achieve high predictive accuracy, but their complex internal architecture hampers direct interpretation.
By contrast, simple models such as multiple linear regression may have lower predictive performance, but are generally considered more interpretable.
However, this simple trade-off picture has been critically challenged.  
Domain expert knowledge can in fact bring the performance of linear models to par with deep neural networks even in challenging data-driven physics problems~\cite{boattini2021averaging, Alkemade_Boattini_Filion_Smallenburg_2022, Alkemade_Smallenburg_Filion_2023}.
Moreover, interpretability is increasingly understood as a context-dependent and human-centered notion, rather than an intrinsic property of a model~\cite{lipton2018mythos, rudin2019stop, murdoch2019definitions}.
Motivated by this perspective, the present paper theoretically examines the difficulty of interpreting the results of multiple linear regression in the context of physics research, where stringent requirements on data-driven models apply~\cite{Wetzel_Ha_Iten_Klopotek_Liu_2025, sharma2026interpretability}. 

First of all, why is multiple linear regression usually believed to be interpretable? 
The reason is its simple additive structure. 
In multiple linear regression, the target variable ${\bm Y}$ is predicted from a set of $d$ features, ${\bm X}^{(1)}, {\bm X}^{(2)}, \ldots, {\bm X}^{(d)}$, as
\begin{equation}
\hat{\bm Y} = \hat w^{(1)} {\bm X}^{(1)} + \hat  w^{(2)} {\bm X}^{(2)} + \cdots + \hat w^{(d)} {\bm X}^{(d)} ,
\label{eq:linear_model}
\end{equation}
where $\hat w^{(f)}$ for $f=1, 2,...,d$ are the weights associated with each feature, and $\hat{\bm Y}$ denotes the prediction of ${\bm Y}$ (which is centered and normalized here). Supervised learning amounts to estimating these weights by minimizing a loss function, such as the mean squared error, 
computed on a training dataset from laboratory experiments or numerical simulations. When ${\bm X}^{(f)}$ for $f=1, 2,...,d$ are normalized, the simple functional form in Eq.~\eqref{eq:linear_model} allows us to interpret the weight values as contributions of each feature to the target variable ${\bm Y}$. 
For example, a large positive value of $\hat{w}^{(f)}$ suggests that increasing $X^{(f)}$ tends to increase the predicted value $\hat{\bm Y}$, whereas a large negative value does the opposite. 
Thus, $w^{(f)}$ provides a simple measure of feature importance~\cite{fisher2019all}.
This apparent one-to-one correspondence between features and weights is the main reason why linear models are often regarded as interpretable.
The wording \emph{multiple linear regression} highlights the presence of multiple input features ($d$ features in our case). Since this point is clear from the context, we will refer to it simply as \emph{linear regression} in the following.

To discuss the physical interpretation of linear regression, we first clarify the conditions under which the fitted weights can be meaningfully interpreted in the context of physics research.
In our view, such an interpretation requires the following two conditions~\cite{sharma2026interpretability}:
\begin{enumerate}
    \item {\it Statistical robustness of the weights}: First, the estimated weights should be robust and reproducible. 
Because the weights are interpreted as measures of feature importance, their values should not depend sensitively on the particular dataset used for training, provided that the datasets are generated under the same physical conditions and sampled from the same underlying distribution. In other words, the variance of the estimated weights across independently sampled training datasets should be small.
    \item {\it Physical consistency of the weights}: Even if the weights are estimated robustly, their signs and magnitudes should be physically meaningful. 
For example, physically similar features should have weights with the same sign and comparable magnitudes. 
They should not appear with largely distinct weights of opposite signs. 
Moreover, extremely large or diverging weights are incompatible with a meaningful physical interpretation.
\end{enumerate}

The first condition concerns dataset-to-dataset fluctuations of the estimated weights. 
This issue is particularly relevant when the dataset size is limited, as is often the case in laboratory experiments or computationally expensive simulations. 
By contrast, the second condition concerns the physical consistency of the weights obtained from a single dataset, even when the dataset is sufficiently large. 
A typical manifestation of this problem is the oscillatory behavior of weights across ordered features. 
Such behavior appears, for example, in spectroscopy regression, where the input features correspond to signal intensities at neighboring frequencies or wavelengths~\cite{brown2009critical, gowen2011preventing},
or in functional regression~\cite{cardot2003spline,james2009functional}, where the features are naturally ordered or continuously indexed, for example by time.
In such cases, physically similar features may be assigned weights with alternating signs and large magnitudes, making direct physical interpretation difficult.
These issues are a consequence of \textit{multicollinearity}, \textit{i.e.}, strong correlations between input features, which frequently arises in high-dimensional feature spaces~\cite{james2013introduction, montgomery2021introduction}.


While dataset-to-dataset fluctuations of estimated weights have been extensively studied in statistical learning, weight oscillations have mainly been discussed in specific domains~\cite{brown2009critical, gowen2011preventing, cardot2003spline, james2009functional} 
To the best of our knowledge, however, the latter issue has often been treated phenomenologically, with the main aim of constructing remedies, such as additional penalty terms, to suppress oscillatory regression coefficients.
Moreover, in some studies, oscillatory regression coefficients are used as indicators of model complexity or overfitting~\cite{gowen2011preventing,takahama2015model}.
However, such oscillations can arise from multicollinearity itself, even in regimes where prediction performance generalizes well and overfitting is absent.

In this paper, we provide a systematic theoretical analysis of how multicollinearity induces both weight fluctuations and weight oscillations.
We analyze these effects from a unified perspective based on the eigenmodes of the feature correlation matrix, for both ordinary least square (OLS) and Ridge regression.
We further argue that the conventional choice of the Ridge regularization parameter, based mainly on reducing overfitting or optimizing predictive performance, is not necessarily optimal for interpretation. Indeed, even when the prediction performance is nearly unchanged, the weight pattern can vary sensitively with the strength of regularization, undermining physical interpretation.
Our analysis is primarily based on the traditional frequentist statistical framework to assess fluctuations in weight estimation, but we also adopt a Bayesian perspective, which offers a complementary way to quantify uncertainty and provides additional insight into the instabilities caused by multicollinearity.
We test our theoretical arguments in detail on two physics datasets: superconductivity data and glassy-dynamics data. To assess the generality of our findings, we also analyze a broader range of datasets beyond physics. 
Our study therefore provides a theoretical explanation for why physical
interpretation can remain difficult even in linear regression when strongly
correlated features are used. Based on these findings, we will finally discuss possible
routes toward retaining interpretability.


The paper is organized as follows.
In Sec.~\ref{sec:linear_regression}, we introduce the linear regression framework and demonstrate, using physics datasets, that fitted weights can exhibit large dataset-to-dataset fluctuations and oscillatory patterns across physically similar features.
These observations motivate the central question addressed in this paper.
In Sec.~\ref{sec:weight_fluctuations}, we analyze dataset-to-dataset weight fluctuations from both theoretical and numerical perspectives.
In Sec.~\ref{sec:weight_oscillation}, we then examine the origin of weight oscillations and clarify their connection to the eigenmodes of the feature correlation matrix.
Finally, in Sec.~\ref{sec:conclusion}, we conclude and discuss future
perspectives on how to retain interpretability.

\section{Linear regression analysis}
\label{sec:linear_regression}

\subsection{Dataset}
\label{sec:datasets}

We mainly consider two regression datasets from physics: a superconductivity
dataset and a glassy-dynamics dataset. To assess the generality of our
theoretical predictions, we also analyze a broader collection of publicly
available datasets. 

\vspace{0.5cm}
\noindent
{\bf Superconductivity dataset.}
Superconductivity is a state in which a material exhibits zero electrical resistance below a material-dependent critical temperature, $T_{\rm c}$. Predicting this critical temperature is an important problem in physics and materials science, because it can steer the search for high-temperature superconducting materials.

We employ the dataset studied in Ref.~\cite{hamidieh2018data}, which was originally compiled from experimental measurements reported in the SuperCon/NIMS database.
Each data point corresponds to one superconducting material (e.g., $\mathrm{Ba}_{0.2}\mathrm{La}_{1.8}\mathrm{CuO}_{4}$, $\mathrm{Nb}_{0.8}\mathrm{Pd}_{0.2}$), and the target variable is the superconducting critical temperature $T_{\rm c}$. The input features are constructed from the chemical composition of the material. More specifically, the dataset contains statistical summaries of eight elemental physical quantities: atomic mass $m$, atomic radius $r$, density $\rho$, valence $\nu$, first ionization energy $\varphi$, electron affinity $\chi$, heat of fusion $H_f$, and thermal conductivity $\kappa$. For each physical quantity, ten statistical descriptors are computed from the constituent elements and their stoichiometric fractions: mean, geometric mean, weighted mean, weighted geometric mean, entropy, weighted entropy, standard deviation, weighted standard deviation, range, and weighted range. We manually arrange the statistical descriptors in an order that reflects our physical intuition. In addition, the number of elements $N$ in the compound is included as one feature. Therefore, the total number of features is $d = 8 \times 10 + 1 = 81$. The full dataset contains $N_{\rm data}=21263$ data points.

In this dataset, physical interpretation through linear regression amounts to asking which features contribute most strongly to the prediction of the critical temperature $T_{\rm c}$.
More specifically, one would like to infer from the fitted weights which elemental physical quantities, and which statistical descriptors of these quantities, are most relevant for predicting $T_{\rm c}$.

\vspace{0.5cm}
\noindent
{\bf Glassy-dynamics dataset.}
The glassy-dynamics dataset concerns the prediction of particle (or atomic) mobility in a glass-forming liquid obtained from computer simulations~\cite{sharma2026interpretability}. 
Glass-forming liquids provide an interesting example of systems in which the relationship between structure and dynamics is highly nontrivial. Although static particle configurations often appear spatially uniform and disordered, their future dynamics can be strongly heterogeneous. This phenomenon is known as dynamic heterogeneity~\cite{berthier2011dynamical,karmakar2014growing}. A central question is to what extent the future mobility of each particle can be predicted from the structural information contained in a static snapshot, see Ref.~\cite{jung2025roadmap} for a review.

In this dataset, each data point corresponds to a particle in a simulated configuration. The target variable is the particle mobility, which quantifies how much particle $i$ moves on average over a prescribed time interval. The input feature vector consists of six physically motivated quantities characterizing the local environment around particle $i$: the packing-efficiency parameter $\overline \Theta$, energy $\overline u$, density $\overline \rho$, orientational order parameter $\overline \Psi_6$, volume fraction $\overline \varphi$, and coordination number $\overline z$. Each physical quantity is systematically spatially coarse-grained over ten different coarse-graining length scales, thereby taking into account structural correlations from short to long distances. Thus, the total number of features is $d = 6 \times 10 = 60$. The full dataset contains $N_{\rm data}=40000$ particle-level data points.

For this dataset, physical interpretability amounts to asking which physical quantity, and which coarse-graining length scale, contributes most strongly to the prediction of future dynamics (mobility). In a linear model, such contributions are reflected in the magnitudes of the corresponding weights. Moreover, positive and negative weights indicate features that enhance and suppress particle mobility, respectively.

\vspace{0.5cm}
\noindent
{\bf Other datasets.} We also consider
publicly available supervised-learning datasets for wine-quality data~\cite{Wine_paper,wine_dataset}, infrared-thermography temperature data~\cite{infrared_paper,infrared_dataset},
appliances energy prediction data~\cite{appliance_energy_paper,appliance_energy_data}, online-news popularity data~\cite{online_news_paper,online_news_data}, and El~Ni\~no data~\cite{el_nino_122}.
The El Ni\~no dataset contains only input features and does not provide a predefined target label.

\subsection{Notation}
\label{sec:dataset}


We introduce the notation used to describe the datasets. We denote by
$\mathcal{D}=\left\{ \left({\bm X}_i, Y_i\right) \right\}_{i=1}^{N_{\rm data}}$ the full dataset.
${\bm X}_i \in \mathbb{R}^d$ is a $d$-dimensional feature vector and
$Y_i \in \mathbb{R}$ is the target variable, and $N_{\rm data}$ is the total number of data points.
To describe the training, validation, and test datasets using a common notation,
we denote by $\mathcal{S}$ a generic subset of the full dataset, such as
$\mathcal{S}_{\rm train}$, $\mathcal{S}_{\rm val}$, or
$\mathcal{S}_{\rm test}$. The number of data points in this subset is
denoted by $N_{\mathcal{S}} = |\mathcal{S}|$.

As is usual in machine learning studies, we normalize the target variable $Y_i$ so that it has zero mean and unit standard deviation.
We then introduce
the target vector
\begin{equation} {\bm Y} = \left[ Y_1, Y_2, \dots, Y_{N_\mathcal{S}} \right]^T \in \mathbb{R}^{N_{\mathcal{S}}}, \label{eq:Y_vector} \end{equation}
where the superscript $T$ is the transpose operation.

Each input feature is also normalized to have zero mean and unit variance, forming the vector ${\bm X}_i$ for data point $i$:
\begin{equation}
{\bm X}_i = \left[ X_i^{(1)}, \ X_i^{(2)}, \ \cdots, \ X_i^{(d)} \right]^T \in \mathbb{R}^d.
\label{eq:feature_X_original}    
\end{equation}
We note that as standard practice in machine learning, when fitting a model, the normalization parameters are computed only from the training dataset, $\mathcal{S}_{\rm train}$.
The features in the validation and test datasets, $\mathcal{S}_{\rm val}$ and $\mathcal{S}_{\rm test}$, are then normalized using the same mean and standard deviation computed from $\mathcal{S}_{\rm train}$.
In all other analyses, where no explicit model fitting is involved, the whole dataset is normalized together.

It is also convenient to introduce the ($N_\mathcal{S} \times d$) design matrix,
\begin{equation}
\mathrm{X} =
\begin{bmatrix}
X_1^{(1)} & X_1^{(2)} & \cdots & X_1^{(d)} \\
X_2^{(1)} & X_2^{(2)} & \cdots & X_2^{(d)} \\
\vdots  & \vdots  & \ddots & \vdots  \\
X_{N_\mathcal{S}}^{(1)} & X_{N_\mathcal{S}}^{(2)} & \cdots & X_{N_\mathcal{S}}^{(d)} 
\end{bmatrix} \in \mathbb{R}^{N_{\mathcal{S}} \times d} .
\label{eq:design_matrix}
\end{equation}
Using Eq.~\eqref{eq:feature_X_original}, $\mathrm{X}$ can be written as 
\begin{equation} \mathrm{X} = \left[ {\bm X}_1, {\bm X}_2,...,{\bm X}_{N_\mathcal{S}} \right]^T . 
\end{equation} 
We also use the column vectors ${\bm X}^{(f)}$ (for $f=1,2,...,d$) of $\mathrm{X}$: \begin{equation} 
\mathrm{X}=[{\bm X}^{(1)}, {\bm X}^{(2)}, ..., {\bm X}^{(d)}] . 
\end{equation} 
The row and column vectors can be distinguished by the subscript and superscript.

\subsection{Correlation measures}

In this paper, we use two correlation measures: the Pearson correlation coefficient $R$ and the coefficient of determination $R^2$.

\begin{figure}[!htp]
\includegraphics[width=\columnwidth]{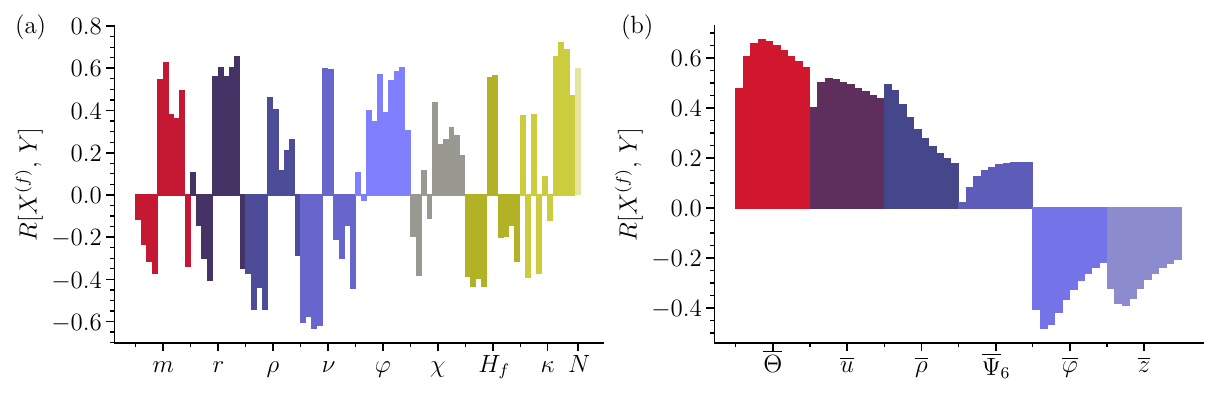}
\caption{
Pearson correlation coefficient, $\R{X^{(f)}}{Y}$,
between the target variable $Y$ and each feature
$X^{(f)}$ for $f=1,\ldots,d$, for the superconductivity dataset
(a) and the glassy-dynamics dataset (b).}
\label{fig:pearson}
\end{figure}

To evaluate the linear dependence between two variables, $A$ and $B$, computed for a subset $\mathcal{S}$ of data points, we use the Pearson correlation coefficient,
\begin{equation}
    \R{A}{B} = \frac{1}{N_\mathcal{S}} \sum_{i\in \mathcal{S}} \frac{(A_i - \mean{A})(B_i - \mean{B})}{\sqrt{\mathrm{Var}[A]\ \mathrm{Var}[B]}} ,
    \label{eq:pearson}
\end{equation}
where $\mean{(\cdots)}$ and $\mathrm{Var}[(\cdots)]$ are the mean and variance, respectively.  
By construction, $\R{A}{B}$ takes values in the range $-1 \leq \R{A}{B} \leq 1$.  

We first consider the Pearson coefficients $\R{X^{(f)}}{Y}$ between the target variable $Y$ and each feature $X^{(f)}$.
The Pearson coefficients are conveniently assembled in vector form
\begin{equation}
    \Rb{X}{Y} = \left[ \R{X^{(1)}}{Y}, \R{X^{(2)}}{Y}, \  \dots, \ \R{X^{(d)}}{Y} \right]^T  \in \mathbb{R}^d.
    \label{eq:Pearson_Y_X_vector}
\end{equation}
We show $\R{X^{(f)}}{Y}$ in Fig.~\ref{fig:pearson} for (a) the
superconductivity dataset and (b) the glassy-dynamics dataset.
The absolute value of $R$ quantifies the linear correlation between each feature
$X^{(f)}$ and the target variable $Y$. 
Thus, this quantity provides a simple preliminary way to assess the relevance
of each feature before fitting a machine-learning model.

\begin{figure}[!htp]
\includegraphics[width=\columnwidth]{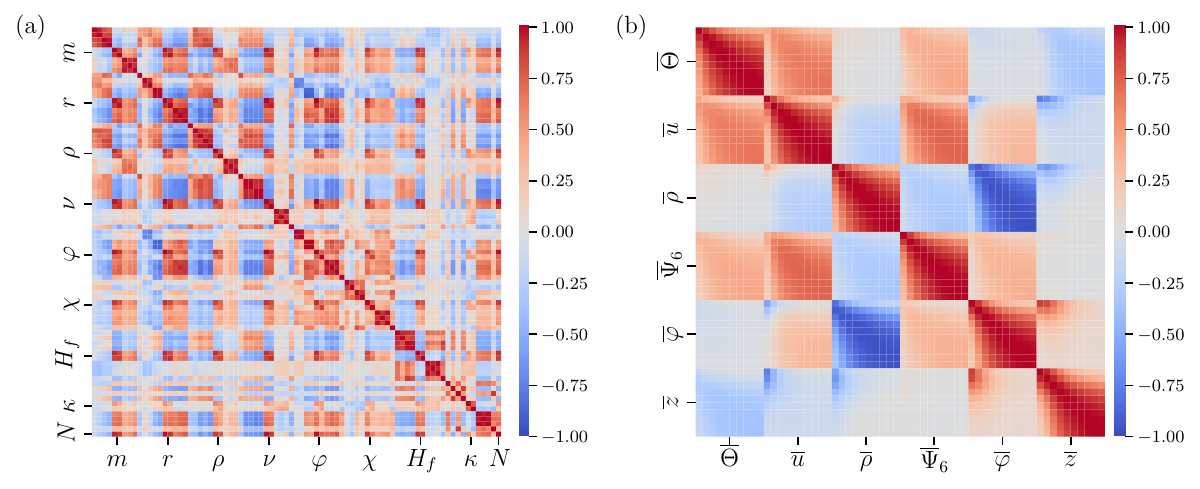}
\caption{
Feature correlation matrix $\mathrm{C}$ for the superconductivity dataset
(a) and the glassy-dynamics dataset (b). }
\label{fig:heatmaps}
\end{figure}

To quantify cross-correlations between groups of features, we introduce a correlation matrix, $\mathrm{C}$, whose elements are
\begin{equation}
    \mathrm{C}_{f, f'} = \R{X^{(f)}}{X^{(f')}} .
    \label{eq:correlation_matrix}
\end{equation}
In Fig.~\ref{fig:heatmaps}, we show the feature correlation matrix
for (a) the superconductivity dataset and (b) the glassy-dynamics dataset.
In both datasets, we find blocks of highly correlated (or anti-correlated) features, shown as red (blue) regions,
particularly along the diagonal of the correlation matrix.
In the superconductivity dataset, they
mainly correspond to similar statistical descriptors constructed from the same
elemental physical quantity. In the glassy-dynamics dataset, they correspond to
coarse-grained features obtained from the same physical quantity at different
coarse-graining length scales.

In the following, we will use the Pearson correlation coefficient also to evaluate
regression performance. Specifically, we compute the correlation between the
ground-truth target values $\bm{Y}$ and the predicted values $\hat{\bm{Y}}$,
namely $\R{Y}{\hat{Y}}$.
In addition, we also use the coefficient of determination $R^2$, which is another standard measure of regression performance.
If $A$ is the variable to predict, $R^2[A, \hat A]$ is given by
\begin{equation}
    R^2[A, \hat A] = 1 - \frac{\sum_{i \in \mathcal{S}} \left(A_i - \hat{A}_i\right)^2}{\sum_{i \in \mathcal{S}} \left(A_i - \mean{A} \right)^2} = 1 - \frac{\| {\bm A} - \hat {\bm A} \|^2}{N_\mathcal{S} {\rm Var}[A]}.
    \label{eq:coefficient_determination}
\end{equation}
When the prediction is perfect, $R^2[A, \hat A] = 1$.
Conversely, when the prediction always outputs the mean value, $R^2[A, \hat A] = 0$, which serves as a baseline.
Note that $1-R^2[A, \hat A]$ is proportional to the mean-squared-error, with the proportionality factor given by the variance of the data.

\subsection{Eigenmode decomposition}

It is useful to consider an eigenmode decomposition of the correlation matrix to analyze and manage the data. 
The matrix $\mathrm{C}$ can be diagonalized using the orthogonal matrix  
\[
\mathrm{U} = \big[{\bm u}^{(1)}, {\bm u}^{(2)}, \dots, {\bm u}^{(d)}\big] \in \mathbb{R}^{d \times d}, 
\]  
where ${\bm u}^{(k)} \in \mathbb{R}^d$ are the eigenvectors of $\mathrm{C}$. This gives the eigenvalue decomposition of $\mathrm{C}$,  
\begin{equation}
    \mathrm{C} = \mathrm{U} \mathrm{\Lambda} \mathrm{U}^T 
    = \sum_{k=1}^d \lambda^{(k)}\,{\bm u}^{(k)} {{\bm u}^{(k)}}^T,
        \label{eq:eigen_decomposition}
\end{equation}  
where $\mathrm{\Lambda} \in \mathbb{R}^{d \times d}$ is a diagonal matrix with eigenvalues 
$\lambda^{(1)}, \dots, \lambda^{(d)}$, sorted as 
$\lambda_{\rm max}=\lambda^{(1)} \geq \dots \geq \lambda^{(d)}=\lambda_{\rm min} > 0$.  
Similarly, the eigenvalue decomposition of the inverse matrix, $\mathrm{C}^{-1}$, is given by  
\begin{equation}
    \mathrm{C}^{-1} = \mathrm{U} \mathrm{\Lambda}^{-1} \mathrm{U}^T 
    = \sum_{k=1}^d \frac{1}{\lambda^{(k)}}\,{\bm u}^{(k)} {{\bm u}^{(k)}}^T.
    \label{eq:eigen_decomposition_inverse}
\end{equation}

\subsection{Linear models}

We consider a simple linear regression model to predict  $Y_i$ based on the feature vector ${\bm X}_i$~\cite{bishop2006pattern}:
\begin{equation}
    \hat Y_i = \hat {\bm w}^T {\bm X}_i = \sum_{f=1}^d \hat w^{(f)} X_i^{(f)} ,
    \label{eq:linear_model_original}
\end{equation}
where $\hat {\bm w} = \left[ \hat w^{(1)}, \hat w^{(2)}, ...,  \hat w^{(d)} \right]^T$ are the weights.
With the vector and matrix notations in Eqs.~\eqref{eq:Y_vector} and \eqref{eq:design_matrix}, Eq.~\eqref{eq:linear_model_original} is rewritten as
\begin{equation}
    \hat {\bm Y} = \mathrm{X}\hat {\bm w} .
    \label{eq:prediction}
\end{equation}
To determine the weights, we minimize a loss function given by
\begin{equation}
  \mathcal{L}({\bm w}) = \mathcal{L}^{\rm MSE}({\bm w}) + \mathcal{L}^{\rm reg}({\bm w}) ,
  \label{eq:loss}
\end{equation}
where
\begin{equation}
  \mathcal{L}^{\rm MSE}({\bm w}) = \frac{1}{2N_\mathcal{S}} \sum_{i \in \mathcal{S}} (\hat{Y}_i - Y_i)^2 = \frac{1}{2 N_\mathcal{S}} ||\hat{\bm Y}-{\bm Y}||^2
  \label{eq:MSE}
\end{equation}
is the mean-squared-error (MSE) and $\mathcal{L}^{\rm reg}$ is a regularization term. 
In this paper, we consider Ridge regression~\cite{bishop2006pattern}, which includes a quadratic penalty term controlled by the regularization hyperparameter $\alpha$,
\begin{equation}
\mathcal{L}^{\rm Ridge}({\bm w}) = \mathcal{L}^{\rm MSE}({\bm w}) + \frac{\alpha}{2} {\bm w}^T {\bm w} .    
\end{equation}
%
With Eq.~\eqref{eq:prediction}, $ \mathcal{L}^{\rm Ridge}({\bm w})$ is rewritten as
\begin{align}
    \mathcal{L}^{\rm Ridge}({\bm w}) = \frac{1}{2 N_\mathcal{S}} \left( {\bm w}^T \mathrm{X}^T \mathrm{X} {\bm w} - 2 \left(\mathrm{X}^T {\bm Y}\right)^T {\bm w} + {\bm Y}^T {\bm Y} \right) \nonumber + \frac{\alpha}{2} {\bm w}^T {\bm w}.
\end{align}
The derivative of $\mathcal{L}^{\rm Ridge}({\bm w})$ with respect to ${\bm w}$ is
\begin{equation}
    \nabla_{{\bm w}} \mathcal{L}^{\rm Ridge}({\bm w}) = \left(\frac{1}{N_\mathcal{S}} \mathrm{X}^T \mathrm{X} + \alpha \mathrm{I}\right) {\bm w} - \frac{1}{N_\mathcal{S}} \mathrm{X}^T {\bm Y},
    \label{eq:derivative_loss}
\end{equation}
where $\mathrm{I}$ is the $d \times d$ identity matrix.
Since $Y_i$ (elements of ${\bm Y}$) and $X_i^{(f)}$ (elements of $\mathrm{X}$) are normalized to have zero mean and unit variance, we can express the terms in Eq.~\eqref{eq:derivative_loss} using the correlation matrix $\mathrm{C}$ defined in Eq.~\eqref{eq:correlation_matrix} and the Pearson coefficients vector ${\bm R}$ in Eq.~(\ref{eq:Pearson_Y_X_vector}),
\begin{eqnarray}
     \mathrm{C} &=& \frac{1}{N_\mathcal{S}} \mathrm{X}^T \mathrm{X} , \\
     \Rb{X}{Y} &=& \left[\R{X^{(1)}}{Y}, \ \ldots, \ \R{X^{(d)}}{Y} \right]^T = \frac{1}{N_\mathcal{S}} \mathrm{X}^T {\bm Y} .  
\end{eqnarray}
Setting $\nabla_{{\bm w}} \mathcal{L}^{\rm Ridge}({\bm w}) = {\bm 0}$, the solution for Ridge regression is obtained as
\begin{equation}
    \hat{\bm w}_{\rm Ridge} = (\mathrm{C} + \alpha \mathrm{I})^{-1} \Rb{X}{Y}.
    \label{eq:Ridge_solution}
\end{equation}
For $\alpha \to 0$, this reduces to the OLS regression solution,
\begin{equation}
    \hat{\bm w}_{\rm OLS} = \mathrm{C}^{-1} \Rb{X}{Y}.
    \label{eq:OLS_solution}
\end{equation}

The invertibility of $\mathrm{C}$, required by the OLS estimator in Eq.~\eqref{eq:OLS_solution}, is tightly connected to the linear dependence of features.
As a limiting case, the columns of the matrix $\mathrm{X} = [{\bm X}^{(1)}, \ldots, {\bm X}^{(d)}]$ are linearly independent if and only if the correlation matrix $\mathrm{C}$ is invertible.
Conversely, when some features are linearly dependent, $\mathrm{C}$ is not invertible, and hence $\hat{\bm w}_{\rm OLS}$ is not uniquely determined.
The hyperparameter $\alpha$ in Eq.~\eqref{eq:Ridge_solution} serves to regularize the singularity of the matrix in Ridge regression.

Finally, it is convenient to express the loss function around its minimum. This corresponds to the quadratic expansion of $\mathcal{L}^{\rm Ridge}({\bm w})$ around $\hat{\bm w}_{\rm Ridge}$, namely,
\begin{eqnarray}
    \mathcal{L}^{\rm Ridge}({\bm w}) &=& \frac{1}{2\NS} \| {\bm Y} - \mathrm{X}{\bm w} \|^2 
    + \frac{\alpha}{2} {\bm w}^T {\bm w} \nonumber \\
    &=&
    \frac{1}{2} ({\bm w}-\hat {\bm w}_{\rm Ridge})^T (\mathrm{C}+\alpha \mathrm{I})({\bm w}-\hat {\bm w}_{\rm Ridge})
      - \frac{1}{2} \hat {\bm w}_{\rm Ridge}^T (\mathrm{C}+\alpha \mathrm{I}) \hat {\bm w}_{\rm Ridge}
      + \frac{1}{2N_\mathcal{S}} \| {\bm Y} \|^2 . \nonumber \\
      \label{eq:Loss_expansion}
\end{eqnarray}

\subsection{Prediction performance}

\begin{figure}
\includegraphics[width=\columnwidth]{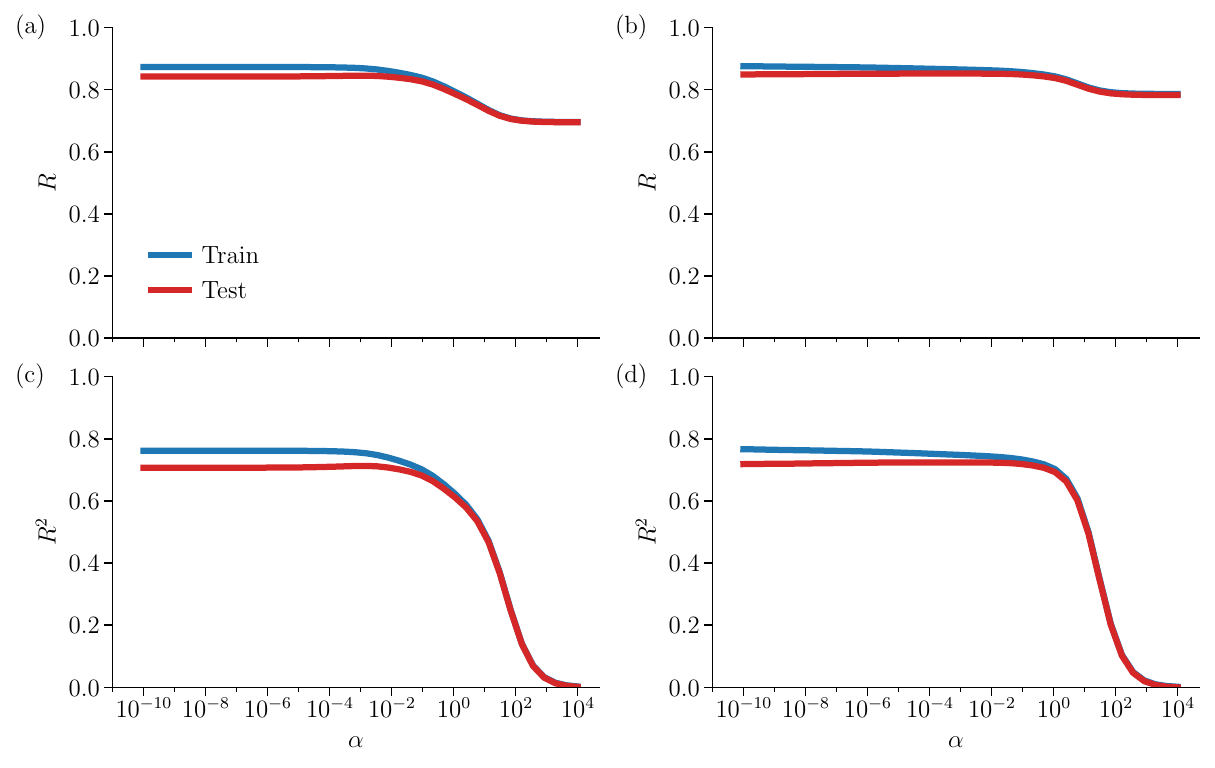}
\caption{(a,b) Pearson correlation coefficient $R$ between the ground truth target values $Y$
and the ridge-regression predictions $\hat Y$ as a function of the regularization
strength $\alpha$, for the superconductivity dataset (a) and the
glassy-dynamics dataset (b).
(c,d) Coefficient of determination $R^2$ for the same Ridge-regression
analysis, for the superconductivity dataset (c) and the glassy-dynamics
dataset (d).}
\label{fig:regression_R_R2}
\end{figure}

We perform linear regression on the superconductivity and glassy-dynamics
datasets. For the training set, we use
$N_{\mathcal{S}_{\rm train}}=810$ data points for the superconductivity
dataset and $N_{\mathcal{S}_{\rm train}}=600$ data points for the
glassy-dynamics dataset. These values are  ten times larger than
the number of features in each dataset, which represents a moderate and
realistic sample size in many practical applications~\cite{green1991many,austin2015number}. 

Figure~\ref{fig:regression_R_R2} shows the prediction performance for both
datasets, quantified by the Pearson correlation coefficient $\R{Y}{\hat{Y}}$
and the coefficient of determination $R^2[Y,\hat{Y}]$. We compare the
performance on the training and test sets as a function of the Ridge
regularization strength $\alpha$. Overall, the prediction performance is good.
For very small $\alpha$, however, we observe a slight degree of overfitting,
as indicated by the gap between the training and test performances. As
$\alpha$ is increased, the test performance slightly improves, while the
training performance remains essentially unchanged. Around
$\alpha\simeq 10^{-2}-10^{-1}$, the training and test performances nearly coincide,
indicating good generalization. For larger values of $\alpha$, both
$\R{Y}{\hat{Y}}$ and $R^2[Y,\hat{Y}]$ decrease significantly, as also observed in the glassy-dynamics dataset with a different set of input features~\cite{sharma2026interpretability}. 
We observe the same qualitative trends in the other regression datasets we analyzed (not shown). 

The crossover value, $\alpha \simeq 0.1$, above which the prediction performance starts to decrease, can be understood from the Ridge-regression solution in Eq.~\eqref{eq:Ridge_solution}. Ridge
regularization modifies the correlation matrix as
$\mathrm{C}\to \mathrm{C}+\alpha \mathrm{I}$. Since the elements of the
correlation matrix satisfy
$-1 \leq \mathrm{C}_{f,f'} \leq 1$, with diagonal elements
$\mathrm{C}_{f,f}=1$, the value $\alpha\simeq 0.1$ corresponds to a moderate
regularization scale. If $\alpha$ is much larger than this value, the
regularization term dominates the regression, the weights are strongly
shrunk toward zero, and the information contained in $\mathrm{C}$ is
suppressed. This leads to underfitting and therefore to a significant
decrease in prediction performance. Thus, in the present datasets,
$\alpha\simeq 0.1$ provides a good compromise between maintaining high
prediction performance and achieving good generalization.

\subsection{Weight fluctuations and oscillations}

\begin{figure}[t]
\includegraphics[width=\columnwidth]{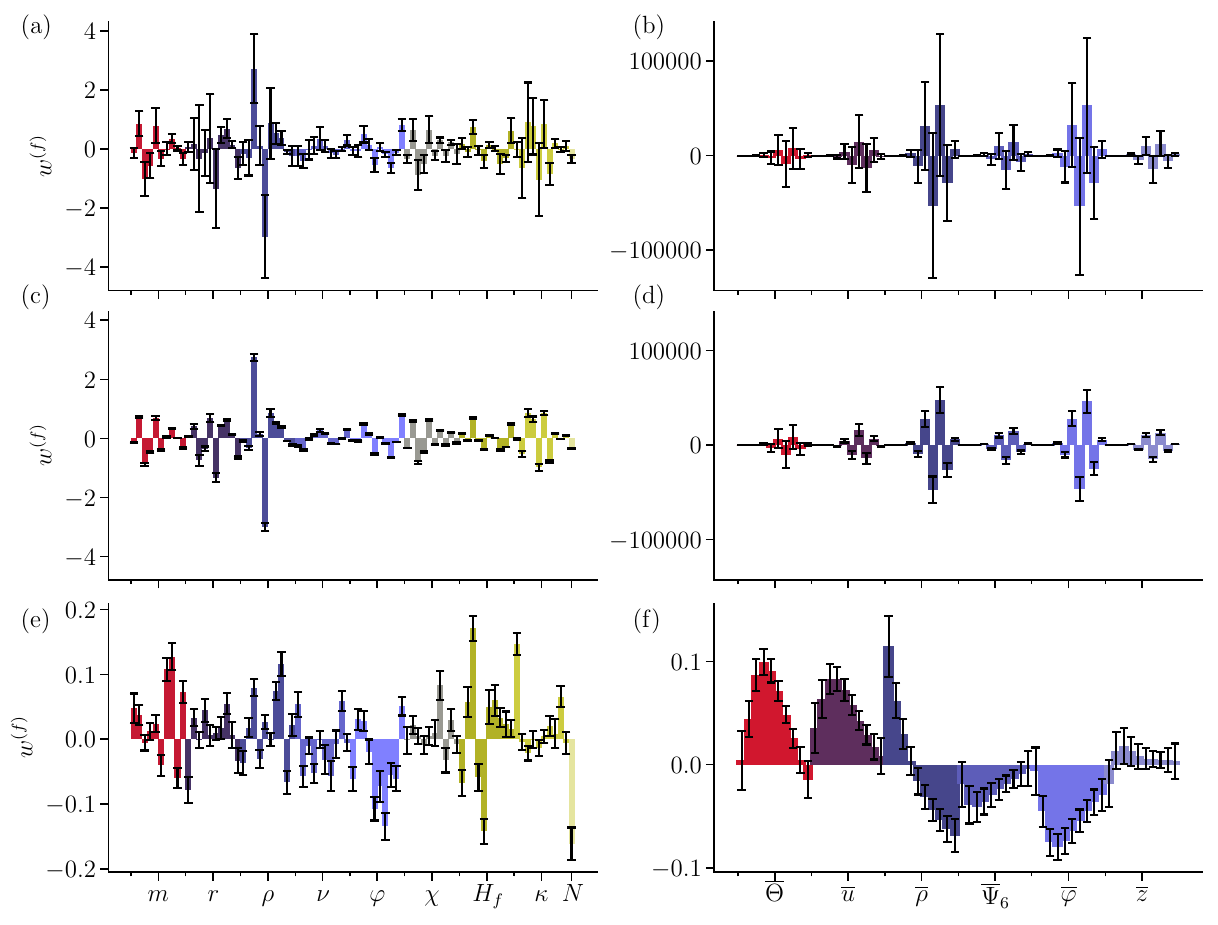}
\caption{(a,b) Average weights obtained by ordinary least-squares (OLS) regression,
averaged over 500 bootstrap realizations with sample size
$N_{\mathcal{S}} = 10d$, plotted as a function of the feature index,
for the superconductivity dataset (a) and the glassy-dynamics dataset (b).
The error bars indicate the standard deviation arising from
dataset-to-dataset fluctuations.
(c,d) The same analysis for the larger sample size
$N_{\mathcal{S}} = 200d$ for the superconductivity dataset (c) and
the glassy-dynamics dataset (d).
(e,f) The same as in (a,b), but for Ridge regression with regularization
strength $\alpha=0.1$, using sample size
$N_{\mathcal{S}} = 10d$, for the superconductivity dataset (e) and
the glassy-dynamics dataset (f).
}
\label{fig:weights}
\end{figure}

Having established that linear models achieve good prediction performance,
we now visualize the fitted weights. Figure~\ref{fig:weights} shows the
weights obtained from ordinary least-squares (OLS) regression as a function
of the feature index, for the superconductivity dataset in panel~(a) and the
glassy-dynamics dataset in panel~(b). The weights are computed over 500
bootstrap realizations with the same sample size,
$N_{\mathcal{S}}= 10d$. We plot the mean weight together with error
bars representing the standard deviation across bootstrap realizations.

We observe large dataset-to-dataset fluctuations in the estimated weights, as
indicated by error bars comparable to, or even larger than, the mean values in
both datasets. Since in practice the model is often fitted to only one training
dataset, this instability directly limits the reliability of weight-based
interpretation. Thus, even when the prediction performance is good, the fitted
weights may not provide a robust measure of feature importance.

We next perform the same analysis (OLS regression) with a significantly larger sample size,
$N_{\mathcal{S}}= 200 d$. Figure~\ref{fig:weights} shows the
corresponding results for the superconductivity dataset in panel~(c) and
the glassy-dynamics dataset in panel~(d). The mean values of the weights are
almost unchanged compared with those obtained for the smaller sample size in
panels~(a,b), whereas the error bars are substantially reduced, as expected.
For this large sample size, we obtain
$R^2[Y,\hat{Y}]=0.738$ and $0.735$ on the training and test sets, respectively,
for the superconductivity dataset, and
$R^2[Y,\hat{Y}]=0.741$ and $0.739$ on the training and test sets, respectively,
for the glassy-dynamics dataset. These results indicate good generalization
without regularization.
We note that ordinary least-squares (OLS) regression remains widely used in practical applications, including physics and materials science, owing to its simplicity and transparency, particularly when the sample size is sufficiently large~\cite{lee2016prediction,seko2014machine,fransson2020efficient,marchand2023multiscale}.

However, even in this regime of good prediction performance and reduced
statistical fluctuations, a serious problem remains for interpretation.
Namely, the weights exhibit strong oscillations as a function of the feature
index. This means that physically similar features, which are placed next to
each other in the feature vector, can be assigned weights with opposite signs.
For example, in the superconductivity dataset, the mean and geometric mean
of the atomic mass are expected to carry similar physical information. Indeed,
these features are strongly correlated, as shown in the correlation matrix in
Fig.~\ref{fig:heatmaps}(a), and they have Pearson correlations
with the target variable of the same sign, as shown in
Fig.~\ref{fig:pearson}(a). Nevertheless, the corresponding
regression weights have opposite signs. A similar issue occurs in the glassy-dynamics
dataset: features coarse-grained over nearby length scales show pronounced oscillations.
These observations severely hinder the interpretability of the weights in the sense of feature importance. Even when overfitting is absent
and dataset-to-dataset fluctuations are suppressed, one may still have physically inconsistent and oscillatory weights.

Finally, we examine weight fluctuations under Ridge regularization with
$\alpha=0.1$, using the smaller sample size
$N_{\mathcal{S}}= 10d$. As discussed above, this value of $\alpha$
gives good prediction performance. Figures~\ref{fig:weights}(e)
and~\ref{fig:weights}(f) show the corresponding results for the
superconductivity and glassy-dynamics datasets, respectively. Compared with
the OLS results in Figs.~\ref{fig:weights}(a,b), the error bars are
substantially reduced. In addition, the strong oscillations observed in the
OLS weights are largely mitigated. In particular, the Ridge-regression weights
tend to follow the sign pattern of the Pearson correlations
$\R{X^{(f)}}{Y}$ shown in Figs.~\ref{fig:pearson}(a,b).

This behavior can be easily understood from Eq.~\eqref{eq:Ridge_solution}. In the limit of large $\alpha$, the term
$\alpha\mathrm{I}$ dominates the matrix $\mathrm{C}+\alpha\mathrm{I}$, and
the Ridge weights have essentially the same pattern as the
Pearson correlations $\Rb{X}{Y}$, up to an overall suppression factor
$\alpha^{-1}$, i.e., $\hat{\bm w}_{\rm Ridge} \approx \alpha ^{-1} \Rb{X}{Y}$.  Thus, the result for $\alpha=0.1$ can be viewed as an
intermediate regime between the strongly oscillatory OLS weights ($\alpha \to 0$) and the
trivial large-$\alpha$ limit controlled by $\Rb{X}{Y}$.

However, Ridge regularization does not by itself provide
a fundamental justification for interpreting the fitted weights as physical
feature importances. Although Ridge regression stabilizes the weights and
suppresses oscillations, the resulting weight pattern can change drastically with the regularization strength $\alpha$, even when the prediction performance is essentially unchanged~\cite{montgomery2021introduction,sharma2026interpretability}.
This also suggests that the conventional criterion for choosing $\alpha$, based on maximizing the test-performance score, does not necessarily lead to an optimal regularization strength from the viewpoint of interpretation.
Therefore, Ridge weights should be interpreted with caution as well.

With these ideas in mind, we will now proceed to theoretically clarify how multicollinearity can induce both large dataset-to-dataset fluctuations of the fitted weights and oscillatory weight patterns. These two effects will be analyzed from a unified perspective that allows us to disentangle related but distinct phenomena that are often confused in the literature.


\section{Dataset-to-dataset weight fluctuations}
\label{sec:weight_fluctuations}

We have numerically observed that the estimated weight vector 
$\hat{\bm w}$ exhibits considerable fluctuations. 
Our goal in this section is to disentangle the origins of these fluctuations, separating 
the effects of dataset size from those due purely to multicollinearity.
In contrast to standard statistics textbooks~\cite{james2013introduction, montgomery2021introduction}, our theoretical setting does not assume any specific model for connecting $X$ and $Y$, except for the estimators in Eqs.~\eqref{eq:Ridge_solution} and \eqref{eq:OLS_solution}.
We then derive general relations for a standard metric of multicollinearity, namely the variance inflation factor (VIF), and critically examine the role of assumptions typically made in standard treatments.
Our analysis of the fluctuations of the point-estimated weights is based on the frequentist approach, but we will also briefly discuss their interpretation from the Bayesian perspective.

\subsection{Setting}

For simplicity of explanation, we first consider the OLS estimator, and we will generalize our analysis to the Ridge estimator at the end.
We wish to evaluate the fluctuations of the weight 
$\hat{\bm w}_{\rm OLS}$. 
The estimator $\hat{\bm w}_{\rm OLS}$ depends on the dataset 
$(\mathrm{X}, \bm Y)$, and can be written explicitly as
$\hat{\bm w}_{\rm OLS} = (\mathrm{X}^T \mathrm{X})^{-1} \mathrm{X}^T \bm Y$.
In particular, we aim to compute the covariance of 
$\hat{\bm w}_{\rm OLS}$ arising from dataset-to-dataset fluctuations, 
characterized by the joint probability distribution 
$P(\mathrm{X}, \bm Y)$.

By the product rule of probability, $P(\mathrm{X}, \bm Y) = P(\bm Y | \mathrm{X}) P(\mathrm{X})$,
we can decompose the total fluctuations into  
(i) fluctuations due to the randomness of $\bm Y$ given $\mathrm{X}$, 
namely governed by $P(\bm Y | \mathrm{X})$, and  
(ii) fluctuations due to the randomness of $\mathrm{X}$, 
governed by $P(\mathrm{X})$.
In the following, we will use the notation for the mean and conditional mean,
\begin{eqnarray}
    \mathbb{E}[(\cdots)] 
    &=& \int d\mathrm{X} \int d\bm Y \, 
        P(\mathrm{X}, \bm Y)(\cdots), \\
    \mathbb{E}[(\cdots)| \ \mathrm{X}] 
    &=& \int d\bm Y \, P(\bm Y | \mathrm{X})(\cdots).
\end{eqnarray}
Obviously, we have $\mathbb{E}[(\cdots)] 
= \int d\mathrm{X}\, P(\mathrm{X}) 
  \, \mathbb{E}[(\cdots) \mid \mathrm{X}]$.
  Besides, since $\NS$ datapoints are extracted independently and 
identically distributed (i.i.d.), we can write
$P(\mathrm{X}, \bm Y) = \prod_{i=1}^{\NS} P(\bm X_i, Y_i)$, $\int d\mathrm{X} = \prod_{i=1}^{\NS} \int d\bm X_i$, and $\int d\bm Y = \prod_{i=1}^{\NS} \int dY_i $.



First, we compute the fluctuations due to $P(\bm Y | \mathrm{X})$. 
The conditional mean of $\hat{\bm w}_{\rm OLS}$ is given by
\begin{equation}
\mathbb{E}[\hat{\bm w}_{\rm OLS} \mid \mathrm{X}]
= (\mathrm{X}^T \mathrm{X})^{-1}\mathrm{X}^T \, \mathbb{E}[\bm Y \mid \mathrm{X}] .
\label{eq:conditional_mean_w}
\end{equation}
Then the conditional covariance is
\begin{eqnarray}
    \mathrm{Cov}[\hat{\bm w}_{\rm OLS} \mid \mathrm{X}]
    &=& \mathbb{E}\!\left[
        \big(\hat{\bm w}_{\rm OLS} - \mathbb{E}[\hat{\bm w}_{\rm OLS} \mid \mathrm{X}]\big)
        \big(\hat{\bm w}_{\rm OLS} - \mathbb{E}[\hat{\bm w}_{\rm OLS} \mid \mathrm{X}]\big)^T
        \,\middle|\, \mathrm{X} \right] \nonumber \\
    &=& (\mathrm{X}^T \mathrm{X})^{-1} \mathrm{X}^T \,
        \mathbb{E}\!\left[
        (\bm Y - \mathbb{E}[\bm Y \mid \mathrm{X}])
        (\bm Y - \mathbb{E}[\bm Y \mid \mathrm{X}])^T
        \,\middle|\, \mathrm{X} \right]
        \mathrm{X} (\mathrm{X}^T \mathrm{X})^{-1} \nonumber \\
    &=& (\mathrm{X}^T \mathrm{X})^{-1} \mathrm{X}^T \,
        \mathrm{Cov}[\bm Y \mid \mathrm{X}] \,
        \mathrm{X} (\mathrm{X}^T \mathrm{X})^{-1}.
\end{eqnarray}
Since $\NS$ datapoints are sampled i.i.d., we have $\mathrm{Cov}[\bm Y \mid \mathrm{X}]
= \mathrm{Var}[Y \mid \bm X] \, \mathrm{I}$,
where $\mathrm{Var}[Y \mid \bm X] 
= \int dY \, P(Y \mid \bm X)\,(Y - \mathbb{E}[Y \mid \mathrm{X}])^2$. 
Therefore, we obtain
\begin{equation}
    \mathrm{Cov}[\hat{\bm w}_{\rm OLS} \mid \mathrm{X}]
    = \frac{\mathrm{Var}[Y \mid \bm X]}{\NS} \, \mathrm{C}^{-1} .
    \label{eq:cond_cov_w_OLS}
\end{equation}


We then compute the total covariance $\mathrm{Cov}[\hat{\bm w}_{\rm OLS}]$.  
To this end, we use the law of total covariance:
\begin{equation}
    \mathrm{Cov}[\hat{\bm w}_{\rm OLS}]
    = \mathbb{E}\!\left[\mathrm{Cov}[\hat{\bm w}_{\rm OLS}\mid \mathrm{X}]\right]
      + \mathrm{Cov}\!\left[\mathbb{E}[\hat{\bm w}_{\rm OLS}\mid \mathrm{X}]\right].
    \label{eq:total_variance}
\end{equation}
This follows by writing 
$\hat{\bm w}_{\rm OLS}-\mathbb{E}[\hat{\bm w}_{\rm OLS}]
= \big(\hat{\bm w}_{\rm OLS}-\mathbb{E}[\hat{\bm w}_{\rm OLS}\mid \mathrm{X}]\big)
  + \big(\mathbb{E}[\hat{\bm w}_{\rm OLS}\mid \mathrm{X}]
  - \mathbb{E}[\hat{\bm w}_{\rm OLS}]\big)$
and noting that the cross terms vanish after taking expectations.

Equation~\eqref{eq:total_variance} provides the starting point for analyzing the fluctuations of the weights. 
In the following, we will analyze the first and second terms in 
Eq.~\eqref{eq:total_variance} separately. 
The analysis of the first term, 
$\mathbb{E}\!\left[\mathrm{Cov}[\hat{\bm w}_{\rm OLS}\mid \mathrm{X}]\right]$, 
leads to the VIF, which is a standard metric used in traditional frequentist statistics to evaluate the severity of multicollinearity. 
The second term, 
$\mathrm{Cov}\!\left[\mathbb{E}[\hat{\bm w}_{\rm OLS}\mid \mathrm{X}]\right]$, 
will be analyzed under additional assumptions.

\subsection{Variance inflation factor (VIF)}

We now analyze the first term in Eq.~\eqref{eq:total_variance},
\begin{equation}
    \mathbb{E}\!\left[ \mathrm{Cov}[\hat{\bm w}_{\rm OLS}\mid \mathrm{X}] \right] 
    = \int d\mathrm{X}\, P(\mathrm{X})\, 
      \mathrm{Cov}[\hat{\bm w}_{\rm OLS} \mid \mathrm{X}] .
\end{equation}
We will assess how 
$\mathrm{Cov}[\hat{\bm w}_{\rm OLS}\mid \mathrm{X}]$, 
as given in Eq.~\eqref{eq:cond_cov_w_OLS}, behaves in the presence of 
multicollinearity.
Equation~\eqref{eq:cond_cov_w_OLS} shows that the invertibility of the 
correlation matrix $\mathrm{C}$ is directly connected to the properties 
of the covariance of $\hat{\bm w}_{\rm OLS}$, suggesting that the latter 
is sensitive to the singularity of $\mathrm{C}$. 
In particular, we are interested in the variance of the weight 
$\hat w^{(f)}_{\rm OLS}$ associated with each feature, which is given by 
the diagonal element of Eq.~\eqref{eq:cond_cov_w_OLS},
\begin{equation}
    \mathrm{Var}\!\left[\hat w^{(f)}_{\rm OLS} \mid \mathrm{X} \right] 
    = \frac{\mathrm{Var}[Y \mid \bm X]}{N_\mathcal{S}} 
      \left( \mathrm{C}^{-1} \right)_{f, f}.
    \label{eq:variance_wights_ff}
\end{equation}

We now focus on how $\left( \mathrm{C}^{-1} \right)_{f, f}$ is related 
to the linear dependence among features. 
Specifically, we ask whether a given feature ${\bm X}^{(f)}$ can be 
expressed as a linear combination of the other features 
${\bm X}^{(f')}$ for $f' = 1, \dots, d$ with $f' \neq f$. 
To this end, we consider the following OLS regression:
\begin{equation}
    \hat {\bm X}^{(f)} 
    = \sum_{\substack{f' = 1 \\ (f' \neq f)}}^d 
      \hat v_{\rm OLS}^{(f')} {\bm X}^{(f')} ,
    \label{eq:linear_model_hat_X}
\end{equation}
where $\hat v_{\rm OLS}^{(f')}$ are the OLS regression weights.
To quantify the performance of this regression, we employ the coefficient 
of determination $R^2[X^{(f)}, \hat X^{(f)}]$ defined in 
Eq.~\eqref{eq:coefficient_determination}. 
Since a good fit results in $R^2$ close to one, while a poor fit 
leads to $R^2$ much smaller than one, $R^2[X^{(f)}, \hat X^{(f)}]$ provides a quantitative measure of the severity of multicollinearity for feature $f$.

As shown in Appendix~\ref{sec:derivation_VIF}, the diagonal elements of 
$\mathrm{C}^{-1}$ are directly related to 
$R^2[X^{(f)}, \hat X^{(f)}]$, namely,
\[
(\mathrm{C}^{-1})_{f,f} = \frac{1}{1-R^2[X^{(f)}, \hat X^{(f)}]} .
\]
Hence with Eq.~\eqref{eq:variance_wights_ff}, we arrive at
\begin{equation}
    \mathrm{Var}\!\left[\hat w^{(f)}_{\rm OLS} \mid \mathrm{X} \right] 
    = \frac{\mathrm{Var}[Y \mid \bm X]}{N_\mathcal{S}} \,
      \frac{1}{1-R^2[X^{(f)}, \hat X^{(f)}]} .
    \label{eq:variance_R2}
\end{equation}
Equation~\eqref{eq:variance_R2} provides a clear interpretation of when 
the (conditional) variance of the weight can become very large. 
First, $\mathrm{Var}[Y \mid \bm X]$ is finite and can be estimated 
numerically.
As usual, 
$\mathrm{Var}\!\left[\hat w^{(f)}_{\rm OLS} \mid \mathrm{X} \right]$ 
is proportional to $1/N_\mathcal{S}$, hence the variance can be large when 
the number of datapoints $N_\mathcal{S}$ is small. 
In addition, and most importantly, when a feature ${\bm X}^{(f)}$ is strongly linearly 
dependent on the other features (i.e., $R^2[X^{(f)}, \hat X^{(f)}] \to 1$), 
the variance of the corresponding weight $\hat w^{(f)}_{\rm OLS}$ 
diverges. 
Thus, Eq.~\eqref{eq:variance_R2} nicely disentangles the effects of limited 
sample size and multicollinearity. 
This implies that the variance can diverge even if one has a sufficiently 
large dataset, purely because of multicollinearity.

Therefore, one can single out the contribution from 
multicollinearity and introduce a metric to quantify its impact. 
In particular, one can define the VIF,
\begin{equation}
    \mathrm{VIF}^{(f)} = \frac{1}{1-R^2[X^{(f)}, \hat X^{(f)}]} ,
\end{equation}
which provides a quantitative measure of the severity of multicollinearity for each feature~\cite{james2013introduction,montgomery2021introduction}. 
One can use this metric to evaluate the magnitude of multicollinearity 
across features. 
Moreover, a global measure of multicollinearity can be obtained by 
considering, for example, the maximum of all $\mathrm{VIF}^{(f)}$.

\begin{figure}
\includegraphics[width=\columnwidth]{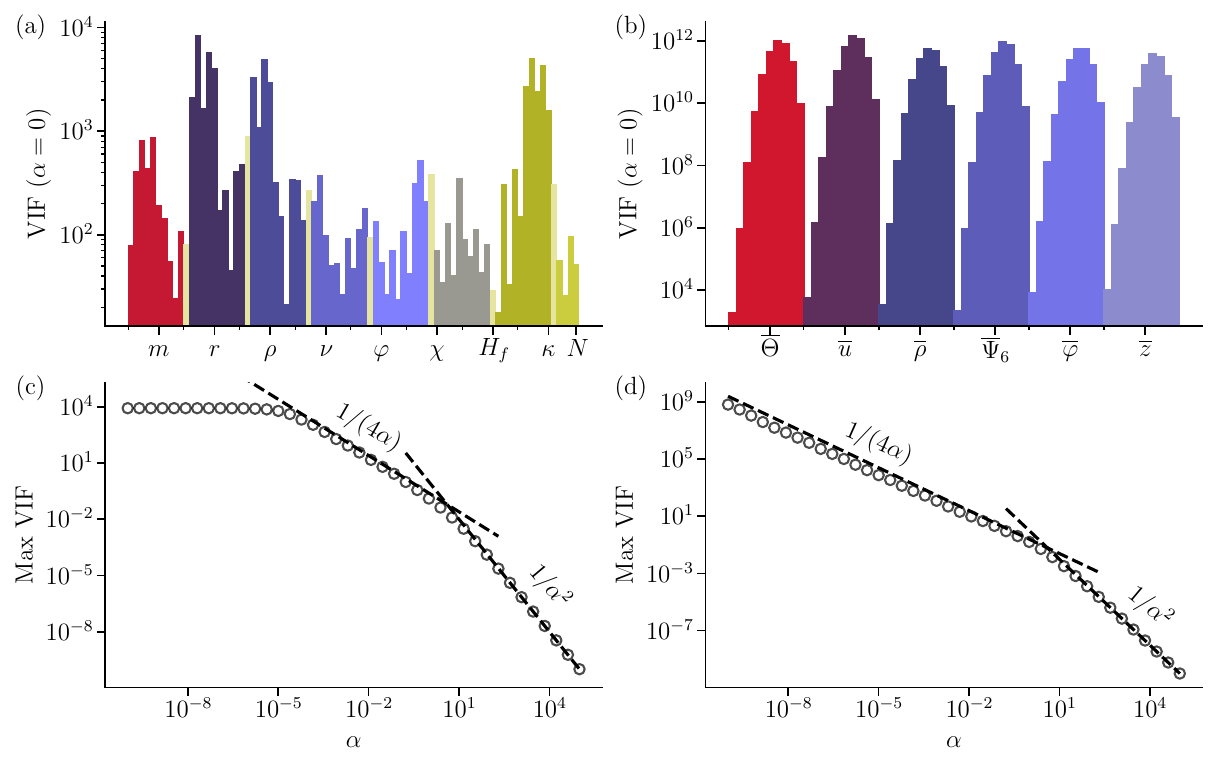}
\caption{(a,b) Variance inflation factor (VIF) for each feature in the OLS setting
($\alpha=0$), for the superconductivity dataset~(a) and the glassy-dynamics
dataset~(b). 
(c,d) Maximum VIF among all features as a function of the Ridge regularization
strength $\alpha$, for the superconductivity dataset~(c) and the
glassy-dynamics dataset~(d).}
\label{fig:VIF_each}
\end{figure}

Figures~\ref{fig:VIF_each}(a,b) show the VIF for each feature in the
superconductivity and glassy-dynamics datasets, respectively. 
As a rule of thumb, a VIF larger than 10 is often taken as an
indication of severe multicollinearity~\cite{montgomery2021introduction}.
In the
superconductivity dataset, the weighted mean of the atomic radius has a very
large VIF around $10^4$, indicating that this feature can be accurately regressed from the
other features and is therefore highly redundant in the linear-feature space.
Nevertheless, this feature has one of the largest OLS weights, as shown in
Figs.~\ref{fig:weights}(a,c). This highlights a central difficulty in
interpreting the weights of linear regression. Namely, a large fitted weight does not
necessarily imply a unique physical contribution, but may instead reflect the
instability caused by multicollinearity.

The glassy-dynamics dataset exhibits even larger VIF values, demonstrating
that multicollinearity is more severe in this dataset. This is consistent with
the descriptor construction, where physically related quantities are
coarse-grained over nearby length scales and therefore become strongly
correlated.

\subsection{Conditional-mean contribution}

We now analyze the conditional-mean contribution, corresponding to the
second term in Eq.~\eqref{eq:total_variance}, which is 
\begin{equation}
    \mathrm{Cov}\!\left[\mathbb{E}[\hat{\bm w}_{\rm OLS}\mid \mathrm{X}]\right] 
    = \int d \mathrm{X}\, P(\mathrm{X}) \,
      \big(\mathbb{E}[\hat{\bm w}_{\rm OLS} \mid \mathrm{X}]
      -\mathbb{E}[\hat{\bm w}_{\rm OLS}]\big)
      \big(\mathbb{E}[\hat{\bm w}_{\rm OLS} \mid \mathrm{X}]
      -\mathbb{E}[\hat{\bm w}_{\rm OLS}]\big)^T .
\end{equation}
As shown in Eq.~\eqref{eq:conditional_mean_w}, 
$\mathbb{E}[\hat{\bm w}_{\rm OLS} \mid \mathrm{X}]$ also contains the 
singular factor $(\mathrm{X}^T\mathrm{X})^{-1}$, 
yet it is not obvious how multicollinearity affects its covariance.
Thus, in the following, we analyze this term under an assumption 
commonly adopted in statistics textbooks~\cite{montgomery2021introduction}. 
Namely, we assume that the data are generated by a ``true'' model,
\begin{equation}
    \bm Y = \mathrm{X} \bm w_* + \bm \epsilon ,
    \label{eq:true_model}
\end{equation}
where $\bm w_*$ is the true weight vector and $\bm \epsilon$ is an error 
term whose statistical properties will be specified below. 
Hence,
\[
\mathbb{E}[\bm Y \mid \mathrm{X}]
= \mathrm{X}\bm w_* + \mathbb{E}[\bm \epsilon \mid \mathrm{X}] .
\]
In this case, Eq.~\eqref{eq:conditional_mean_w} becomes
\begin{equation}
    \mathbb{E}[\hat{\bm w}_{\rm OLS} \mid \mathrm{X}]
    = \bm w_* + (\mathrm{X}^T\mathrm{X})^{-1}\mathrm{X}^T 
      \mathbb{E}[\bm \epsilon \mid \mathrm{X}] .
\end{equation}
If one further assumes that the error $\bm \epsilon$ is independent of 
$\mathrm{X}$, i.e., 
$\mathbb{E}[\bm \epsilon \mid \mathrm{X}] = \bm 0$ 
(exogeneity assumption), then
\[
\mathbb{E}[\hat{\bm w}_{\rm OLS} \mid \mathrm{X}] = \bm w_*,
\]
and consequently
\[
\mathrm{Cov}\!\left[\mathbb{E}[\hat{\bm w}_{\rm OLS}\mid \mathrm{X}]\right] 
= \mathrm{0} .
\]

These assumptions are frequently employed. Indeed, most standard statistics
textbooks start from the linear model in Eq.~\eqref{eq:true_model} and impose
the exogeneity assumption when deriving the variance inflation factor.
However, these assumptions are idealized and need not be satisfied exactly in
real datasets.
In practice, 
$\mathrm{Cov}\!\left[\mathbb{E}[\hat{\bm w}_{\rm OLS}\mid \mathrm{X}]\right]$ 
does not vanish. 

To summarize, the second term in Eq.~\eqref{eq:total_variance} is less
well understood than the first term associated with the VIF. Nevertheless,
because both terms on the right-hand side of Eq.~\eqref{eq:total_variance}
are positive semi-definite, the VIF term provides a lower bound on the total
variance. It therefore offers valuable insight into how multicollinearity can
lead to large fluctuations in the estimated weights.

\subsection{VIF under Ridge regularization}

One can perform a similar covariance analysis for the Ridge estimator 
$\hat{\bm w}_{\rm Ridge}$ with regularization parameter $\alpha$. 
In particular, we obtain the conditional covariance,
\begin{equation}
\mathrm{Cov}[\hat{\bm w}_{\rm Ridge} \mid \mathrm{X}]
= \frac{\mathrm{Var}[Y \mid \bm X]}{N_\mathcal{S}} \,
  (\mathrm{C}+\alpha \mathrm{I})^{-1}\mathrm{C}(\mathrm{C}+\alpha \mathrm{I})^{-1} .
\label{eq:cond_cov_w_Ridge}
\end{equation}
The diagonal elements provide the variance of the estimated weights, 
$\mathrm{Var}\!\left[\hat w^{(f)}_{\rm Ridge}\mid \mathrm{X} \right]$. 
Unlike in the OLS case, however, it is difficult to directly relate the 
variance of the weights to the coefficient of determination, 
as in Eq.~\eqref{eq:variance_R2}. 
Nevertheless, a generalized VIF for Ridge regression can be defined 
\cite{marquardt1970generalized,mcdonald2009ridge}:
\begin{equation}
    \mathrm{VIF}^{(f)} 
    = \left( (\mathrm{C}+\alpha \mathrm{I})^{-1} \, 
              \mathrm{C} \, 
              (\mathrm{C}+\alpha \mathrm{I})^{-1} \right)_{f,f} .
    \label{eq:VIF_Ridge}
\end{equation}
Although the definition of the VIF in Ridge regression remains debated 
\cite{garcia2015collinearity}, Eq.~\eqref{eq:VIF_Ridge} serves as a 
useful indicator of instability in weight estimation.

The $\alpha$–dependence of the VIF, particularly the maximum VIF across all features, offers useful insights and practical guidance for choosing the Ridge regularization strength. In Figs.~\ref{fig:VIF_each}(c,d), we show the maximum of
Eq.~\eqref{eq:VIF_Ridge} over all the features as a function of $\alpha$, for
the superconductivity and glassy-dynamics datasets, respectively. As expected,
$\max_f\{\mathrm{VIF}^{(f)}\}$ decreases monotonically with increasing
$\alpha$. To reduce the VIF below the empirical rule-of-thumb threshold
$\mathrm{VIF}=10$~\cite{montgomery2021introduction}, one needs a regularization strength of the order of
$\alpha\simeq 10^{-2}$--$10^{-1}$.
Interestingly, in both datasets we observe two distinct scaling regimes:
an intermediate regime where the maximum VIF scales approximately as
$\alpha^{-1}$, and a large-$\alpha$ regime where it scales as $\alpha^{-2}$.
In addition, the superconductivity dataset exhibits a plateau at very small
\(\alpha\), corresponding to the OLS-like regime where the regularization is
too weak to affect the weights. The glassy-dynamics dataset also exhibits such
a plateau, but only at much smaller values of \(\alpha\), outside the range
shown in the plot.

We now provide a theoretical argument to explain these scaling behaviors,
as well as the small-$\alpha$ plateau, using the
eigenspace decomposition of the feature correlation matrix.
We rewrite Eq.~\eqref{eq:VIF_Ridge} using the eigenvalue decomposition
\begin{equation}
    \mathrm{VIF}^{(f)}
      = {{\bm e}^{(f)}}^{T}(\mathrm{C}+\alpha \mathrm{I})^{-1}\mathrm{C}(\mathrm{C}+\alpha \mathrm{I})^{-1}{{\bm e}^{(f)}}
      = \sum_{k=1}^{d} \frac{\lambda^{(k)}}{(\lambda^{(k)}+\alpha)^{2}}
        \bigl({{\bm e}^{(f)}}^{T}{\bm u}^{(k)}\bigr)^{2},
    \label{eq:VIF_eigenvalue_decomposition}
\end{equation}
where ${\bm e}^{(f)}$ is the unit vector whose $f$-th component is $1$ and all others are $0$.

In the regime of very small $\alpha$,
$\alpha \ll \lambda_{\rm min}$, the denominator in
Eq.~\eqref{eq:VIF_eigenvalue_decomposition} satisfies
$\lambda^{(k)}+\alpha \simeq \lambda^{(k)}$ for all $k$. Therefore,
$\mathrm{VIF}^{(f)}$ becomes insensitive to $\alpha$ and reduces
to its OLS value. This explains the plateau observed in the superconductivity
dataset in Fig.~\ref{fig:VIF_each}(c) at very small $\alpha$, where $\lambda_{\rm min}=4.96\times10^{-5}$.
In the glassy-dynamics dataset, the smallest eigenvalue is much
smaller, $\lambda_{\rm min}=1.96\times10^{-14}$. As a result, the condition
$\alpha \ll \lambda_{\rm min}$ is not reached within the range of $\alpha$
shown in Fig.~\ref{fig:VIF_each}(d). 

When the regularization is strong, specifically for $\alpha \gg \lambda_{\rm max}$, the matrix
$(\mathrm{C}+\alpha \mathrm{I})$ is dominated by the regularization term and can be approximated as
$\mathrm{C}+\alpha \mathrm{I} \simeq \alpha \mathrm{I}$
since all elements of the correlation matrix $\mathrm{C}$ are of order unity.
This yields
\[
\mathrm{VIF}^{(f)}
\simeq \alpha^{-2} \ \mathrm{C}_{f,f} = \alpha^{-2}, 
\]
because the features are normalized, $\mathrm{C}_{f,f}=1$.
As a consequence, the maximum variance inflation factor obeys the same scaling,
\[
\max_f \{\mathrm{VIF}^{(f)}\} \simeq \alpha^{-2}.
\]
This argument explains the asymptotic behavior observed in the
large-\(\alpha\) regime in Figs.~\ref{fig:VIF_each}(c,d). For the
superconductivity and glassy-dynamics datasets, the largest eigenvalues are
\(\lambda_{\max}=31.5\) and \(\lambda_{\max}=21.3\), respectively. Thus, the
asymptotic \(\alpha^{-2}\) scaling is expected when \(\alpha\) becomes much
larger than these values.

\begin{figure}
\includegraphics[width=\columnwidth]{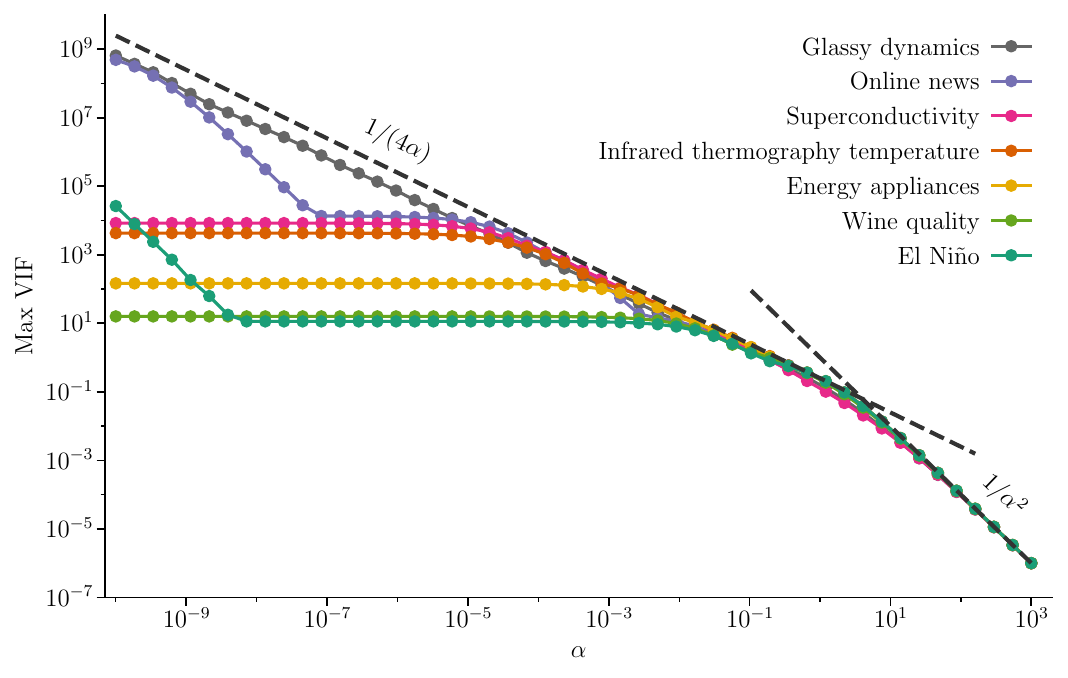}
\caption{Maximum variance inflation factor,
$\max_f\{\mathrm{VIF}^{(f)}\}$, as a function of the ridge regularization
strength $\alpha$ for various datasets, in
addition to the superconductivity and glassy-dynamics datasets.}
\label{fig:VIF_universality}
\end{figure}

Of more practical interest is the intermediate regime $\lambda_{\rm min}\ll \alpha \ll \lambda_{\rm max}$, where the behavior is non-trivial.  
We first define
\[
    g(\lambda)=\frac{\lambda}{(\lambda+\alpha)^{2}}
\quad (\lambda>0,\,\alpha>0),
\]
which attains its maximum at $\lambda_{*}=\alpha$ with $g(\lambda_{*})=1/(4\alpha)$.  
Using $g(\lambda)\le 1/(4\alpha)$ for $\lambda>0$, Eq.~\eqref{eq:VIF_eigenvalue_decomposition} yields, for any $f$,
\begin{equation}
    \mathrm{VIF}^{(f)}
      \le \frac{1}{4\alpha},
      \label{eq:upper_bound}
\end{equation}
where we used completeness of the eigenvectors,
\[
    \sum_{k=1}^{d}\bigl({{\bm e}^{(f)}}^{T}{\bm u}^{(k)}\bigr)^{2}
      ={{\bm e}^{(f)}}^{T}\!\left(\sum_{k=1}^{d}{\bm u}^{(k)}{{\bm u}^{(k)}}^{T}\right)\!{\bm e}^{(f)}
      =1 .
\]
Therefore,
\[
   \max_{f}\{\mathrm{VIF}^{(f)}\} \le \frac{1}{4\alpha},
\]
which provides an upper bound. It indicates how large $\alpha$ must be to reduce the VIF below a desired level.
Indeed, the behavior observed numerically in Figs.~\ref{fig:VIF_each}(c,d)
in the intermediate-$\alpha$ regime is consistent with the bound derived
above.

In Figs.~\ref{fig:VIF_each}(c,d), the superconductivity and glassy-dynamics
datasets nearly saturate the upper bound in Eq.~\eqref{eq:upper_bound}
in the intermediate-$\alpha$ regime. To examine whether this behavior is
accidental or instead represents a generic property of high-dimensional input
features, we computed $\max_f \{\mathrm{VIF}^{(f)}\}$ for several additional
datasets beyond physical sciences, see Sec.~\ref{sec:datasets}. We show the results for all these datasets together in Fig.~\ref{fig:VIF_universality}.
Remarkably, the maximum VIF of all the datasets exhibit essentially the same qualitative trend. At
very small $\alpha$, a plateau appears when $\alpha \ll \lambda_{\rm min}$. This plateau is followed by an intermediate regime in
which the maximum VIF is close to the bound $1/(4\alpha)$. Finally, in the
large-$\alpha$ regime, $\alpha \gg \lambda_{\rm max}$, the maximum VIF decays
as $\alpha^{-2}$.
These empirical observations indicate that the effect of Ridge regularization is largely universal across feature-scaled datasets. The expression $\max_f \{\mathrm{VIF}^{(f)}\}\approx 1/(4\alpha)$ can thus be used as a practical guideline in the intermediate regime. In particular, choosing
$\alpha\simeq 10^{-2}$--$10^{-1}$ is sufficient to reduce the VIF below
the common rule-of-thumb threshold $\mathrm{VIF}=10$.

Some datasets, such as the El~Ni\~no and online-news datasets, show an
additional intermediate plateau. This behavior can be understood as a
consequence of a large spectral gap $\lambda^{(k+1)} \ll \lambda^{(k)}$ in the eigenvalues of the correlation
matrix. When $\alpha$ lies inside such a gap, $\lambda^{(k+1)} \ll \alpha \ll \lambda^{(k)}$,
the set of eigenmodes affected by the regularization remains almost unchanged
over a finite range of $\alpha$, producing a plateau in the maximum VIF.

\subsection{Bayesian perspective}

We have analyzed the fluctuations of the weights from a frequentist 
perspective, namely the dataset-to-dataset fluctuations of the point 
estimator $\hat{\bm w}$. 
We now turn to the Bayesian perspective, where the weights themselves are 
treated as random variables~\cite{bishop2006pattern}. 
For brevity, we focus on Ridge regression, which includes OLS as a 
special case. 
A brief introduction to Bayesian estimation and the derivation of the 
equations used below are provided in Appendix~\ref{app:Bayesian}.

In the Bayesian framework, point estimation of the weights is performed 
via maximization of the conditional probability distribution of the 
weights ${\bm w}$ given the data $(\mathrm{X}, {\bm Y})$, namely 
$P({\bm w}\mid \mathrm{X}, {\bm Y})$, which corresponds to maximum a 
posteriori (MAP) estimation. 
In contrast to the frequentist approach, which focuses only on the point 
estimate, the Bayesian approach retains the full functional form of 
$P({\bm w}\mid \mathrm{X}, {\bm Y})$ as a distribution over ${\bm w}$, 
thereby providing valuable information about uncertainty. 

In a standard setting, the posterior distribution can be expressed as a 
Gaussian expansion around the Ridge estimator
\begin{equation}
    P({\bm w}\mid \mathrm{X}, {\bm Y})
    \propto  \exp \left[ -\frac{1}{2} 
        ({\bm w}-\hat{\bm w}_{\rm Ridge})^T 
        \Sigma_{\rm Ridge}^{-1}
        ({\bm w}-\hat{\bm w}_{\rm Ridge}) \right],
\end{equation}
where $\Sigma_{\rm Ridge}$ is the posterior covariance matrix around the 
Ridge estimator, given by
\begin{equation}
    \Sigma_{\rm Ridge} = \frac{\sigma^2}{N_\mathcal{S}} 
    (\mathrm{C} + \alpha \mathrm{I})^{-1},
\end{equation}
with $\sigma^2$ denoting the error variance, which can be estimated 
separately.

For the OLS case ($\alpha \to 0$), we recover 
$\hat{\bm w}_{\rm Ridge} \to \hat{\bm w}_{\rm OLS}$ and 
$\Sigma_{\rm Ridge} \to \Sigma_{\rm OLS} 
= \tfrac{\sigma^2}{N_\mathcal{S}} \mathrm{C}^{-1}$. 
In this case, the weights ${\bm w}$ exhibit large fluctuations due to the 
singularity of $\mathrm{C}^{-1}$. 
By contrast, when $\alpha>0$, the regularization stabilizes the estimator 
and suppresses these fluctuations, as expected in Ridge regression.

\subsection{Soft modes view}

The Bayesian perspective 
offers an explanation for why the weights can fluctuate strongly while the prediction performance remains good.  
Here, we take the mean-squared-error (MSE) as a performance metric (for the training dataset).
Using Eq.~\eqref{eq:Loss_expansion}, the MSE for the OLS regression can be written as  
\begin{equation}
    \frac{1}{2N_\mathcal{S}} \| {\bm Y} - \mathrm{X}{\bm w} \|^2 
    =
    \frac{1}{2} ({\bm w}-\hat {\bm w}_{\rm OLS})^T \mathrm{C}({\bm w}-\hat {\bm w}_{\rm OLS})
      - \frac{1}{2} \hat {\bm w}_{\rm OLS}^T \mathrm{C} \hat {\bm w}_{\rm OLS}
      + \frac{1}{2N_\mathcal{S}} \| {\bm Y} \|^2 .
      \label{eq:MSE_OLS}
\end{equation}
We then extract the term that depends on ${\bm w}$ and express it using the eigenvalue decomposition in Eq.~\eqref{eq:eigen_decomposition}, 
\begin{equation}
    \frac{1}{2} ({\bm w}-\hat {\bm w}_{\rm OLS})^T \mathrm{C}({\bm w}-\hat {\bm w}_{\rm OLS}) 
    = \frac{1}{2} \sum_{k=1}^d \lambda^{(k)} \big\| {{\bm u}^{(k)}}^T ({\bm w}-\hat {\bm w}_{\rm OLS}) \big\|^2 .
    \label{eq:quadratic_OLS}
\end{equation}
Equations~\eqref{eq:MSE_OLS} and \eqref{eq:quadratic_OLS} demonstrate that ${\bm w}$ can fluctuate significantly along eigenvectors ${\bm u}^{(k)}$ associated with small eigenvalues $\lambda^{(k)}$ without changing the MSE. 
Such directions of the eigenvectors can be regarded as ``soft modes'', 
in analogy with physical systems where low-energy modes correspond to large fluctuations at little cost.

The above theoretical discussion explains why the weight vector $\bm w$ can fluctuate strongly
without substantially affecting the prediction performance, especially in the
presence of strong multicollinearity.
This implies that good prediction performance alone does not guarantee that the estimated weights can be interpreted reliably.

\section{Oscillatory behavior of weights across features}
\label{sec:weight_oscillation}

We now turn our attention to the oscillatory behavior of the weights across features observed in the numerical results. We emphasize that this oscillatory behavior persists even when the dataset size is very large as shown in Fig.~\ref{fig:weights}(c,d). Thus, the phenomenon inherently arises from the structure of a single large dataset. We provide a theoretical argument for how the oscillatory behavior emerges when multicollinearity is severe. 
Again, we first focus on the OLS estimator and later generalize the argument to the case of Ridge regression.

To discuss the oscillatory behavior, it is convenient to use the eigenmodes of the correlation matrix $\mathrm{C}$ in the $d$-dimensional feature space. 
Namely, $\hat {\bm w}_{\rm OLS}$ in Eq.~(\ref{eq:OLS_solution}) is now expressed in view of the eigenvalue decomposition in Eq.~(\ref{eq:eigen_decomposition_inverse}):
\begin{equation}
    \hat {\bm w}_{\rm OLS} = \sum_{k=1}^d \frac{1}{\lambda^{(k)}}\,{\bm u}^{(k)} {{\bm u}^{(k)}}^T \Rb{X}{Y} . 
    \label{eq:OLS_eigenvalue_decomposition}
\end{equation}
This equation provides the central basis for the analysis in this section.

\subsection{Lessons from a two-features model}

To gain some insight into the oscillatory behavior of the weights, it is useful to consider a simple two-features model ($d=2$)~\cite{sharma2026interpretability}, which captures the situation in which two strongly correlated features yield corresponding weights with opposite signs and large magnitudes, as presented in Figs.~\ref{fig:weights}(c,d). 
%
The two-features model is defined by a design matrix 
$\mathrm{X} = [{\bm X}^{(1)}, {\bm X}^{(2)}]$ with the corresponding correlation matrix,  
\begin{equation}
    \mathrm{C} = 
    \begin{bmatrix}
    1 & r \\
    r & 1 
    \end{bmatrix},
\label{eq:C_two_features}
\end{equation}  
where $r$ is the Pearson correlation coefficient between the two features.  
When $r = 0$, the two features are orthogonal, and $\mathrm{C}$ reduces to the identity matrix.  
When $r \to 1$ ($r \to -1$), the two features are perfectly correlated (anti-correlated), and hence ${\bm X}^{(1)}$ and ${\bm X}^{(2)}$ are linearly dependent.  
Here we restrict ourselves to $0 < r < 1$. 

In this setting, the eigenvalues of $\mathrm{C}$ are  
$\lambda^{(1)} = 1+r$ and $\lambda^{(2)} = 1-r$,  
and the corresponding eigenvectors are  
\begin{equation}
    {\bm u}^{(1)} = \frac{1}{\sqrt{2}}[1,1]^T, 
    \qquad 
    {\bm u}^{(2)} = \frac{1}{\sqrt{2}}[1,-1]^T.  
    \label{eq:eigenvectors_two_features}
\end{equation}
The inverse of the correlation matrix is then  
\begin{equation}
    \mathrm{C}^{-1} = \frac{1}{\lambda^{(1)}\lambda^{(2)}}
    \begin{bmatrix}
    1 & -r \\
    -r & 1 
    \end{bmatrix}.
\label{eq:C_inverse_two_features}
\end{equation}
We also assume that $\Rb{X}{Y} = [R^{(1)}, R^{(2)}]^T$ with $R^{(1)} \neq R^{(2)}$.

Using Eq.~\eqref{eq:OLS_eigenvalue_decomposition}, we get
\begin{eqnarray}
  \hat {\bm w}_{\rm OLS} &=&  \frac{1}{\lambda^{(1)}}\,\left( {{\bm u}^{(1)}}^T {\bm R} \right){\bm u}^{(1)}  + \frac{1}{\lambda^{(2)}}\,\left( {{\bm u}^{(2)}}^T {\bm R} \right){\bm u}^{(2)}  \nonumber \\
    &=& \frac{1}{\lambda^{(1)}}\,\left( \frac{R^{(1)}+R^{(2)}}{\sqrt{2}} \right){\bm u}^{(1)}  + \frac{1}{\lambda^{(2)}}\,\left(\frac{R^{(1)}-R^{(2)}}{\sqrt{2}} \right){\bm u}^{(2)} .
    \label{eq:OLS_two_features}
\end{eqnarray}
This equation allows us to extract the essence of the oscillatory behavior. 
First, when multicollinearity is severe ($r \to 1$), the minimum eigenvalue 
$\lambda^{(2)} = 1-r$ vanishes, which leads to the divergence of the second term. 
Moreover, the associated eigenvector ${\bm u}^{(2)}$ in Eq.~\eqref{eq:eigenvectors_two_features} 
corresponds to the anti-symmetric mode, which gives rise to opposite signs for 
$\hat{w}_{\mathrm{OLS}}^{(1)}$ and $\hat{w}_{\mathrm{OLS}}^{(2)}$.

We now argue that the eigenvector associated with the smaller eigenvalue tends to be an anti-symmetric-like mode in general.  
Since $\mathrm{C}{\bm u}^{(k)}=\lambda^{(k)}{\bm u}^{(k)}$ and 
\[
\mathrm{C}=\frac{1}{N_\mathcal{S}}\sum_{i \in \mathcal{S}}{\bm X}_i {{\bm X}_i}^T,
\] 
we have  
\begin{equation}
    \lambda^{(k)} = \frac{1}{N_\mathcal{S}}\sum_{i \in \mathcal{S}}\left( {{\bm u}^{(k)}}^T {\bm X}_i \right)^2 ,
\end{equation}
which demonstrates that $\lambda^{(k)}$ corresponds to the variance of the projected feature ${{\bm u}^{(k)}}^T {\bm X}_i$.  

For the two-features model, we obtain  
\begin{eqnarray}
    {{\bm u}^{(1)}}^T {\bm X}_i &=& \frac{X^{(1)}_i+X^{(2)}_i}{\sqrt{2}}, \\
    {{\bm u}^{(2)}}^T {\bm X}_i &=& \frac{X^{(1)}_i-X^{(2)}_i}{\sqrt{2}} .
\end{eqnarray}
When the two features $X^{(1)}_i$ and $X^{(2)}_i$ are highly correlated, 
the symmetric mode ${\bm u}^{(1)}$ corresponds to the sum, $X^{(1)}_i + X^{(2)}_i$, and thus provides a larger variance (or eigenvalue).  On the other hand, the anti-symmetric mode ${\bm u}^{(2)}$ corresponds to the difference, $X^{(1)}_i - X^{(2)}_i$, which reduces the variance (or eigenvalue).  
We expect this mechanism to be quite generic for the case with $d$ features: whenever two features $X^{(f)}_i$ and $X^{(f')}_i$ (with $f \neq f'$) are highly correlated, they should give rise to an anti-symmetric-like mode with small eigenvalue.

\subsection{Toeplitz matrix approximation}

Moving beyond the simple two-features model, we now discuss the emergence of oscillatory behavior in the presence of multiple correlated features.  

In the correlation matrices shown in Fig.~\ref{fig:heatmaps}, we observed strong feature correlations in the form of block-diagonal structures. A simple way to model these correlation patterns is to approximate each block matrix of size $n \times n$ as a symmetric Toeplitz matrix $\mathrm{T}$, whose elements are given by $\mathrm{T}_{f,f'} = t_{|f-f'|}$ with $f,f'=1,2,\dots,n$.  
The typical block size is $n\simeq 2$--$4$ in the superconductivity
dataset and $n\simeq 10$ in the glassy-dynamics dataset.
In particular, we consider the Kac–Murdock–Szegö (KMS) Toeplitz matrix~\cite{kac1953eigen} defined by  
\[
\mathrm{T}_{f,f'} = r^{|f-f'|},
\]  
where $0<r<1$, and $r$ quantifies the strength of correlation across features.  
The KMS matrix is also known as the correlation matrix of the AR(1) model in statistics, econometrics, and signal processing~\cite{brockwell2002introduction}. 
It has been shown~\cite{kac1953eigen,trench1985eigenvalue,bogoya2016eigenvectors} that the smallest eigenvalue $\lambda_{\mathrm{min}}$ of $\mathrm{T}$ vanishes in the limit $r \to 1$. 
More importantly, the eigenvectors $\bm{u}^{(k)}$ of $\mathrm{T}$ are
sinusoidal-like, with a frequency, or wave number, that increases as the
corresponding eigenvalue $\lambda^{(k)}$ decreases~\cite{bogoya2016eigenvectors}.
In view of Eq.~\eqref{eq:OLS_eigenvalue_decomposition}, we expect a similar mechanism to be at play in OLS: the small-eigenvalue modes amplified in the OLS solution naturally generate oscillatory weight patterns.



\subsection{Weight oscillations under ridge regularization}

\begin{figure}[t]
  \includegraphics[width=\columnwidth]{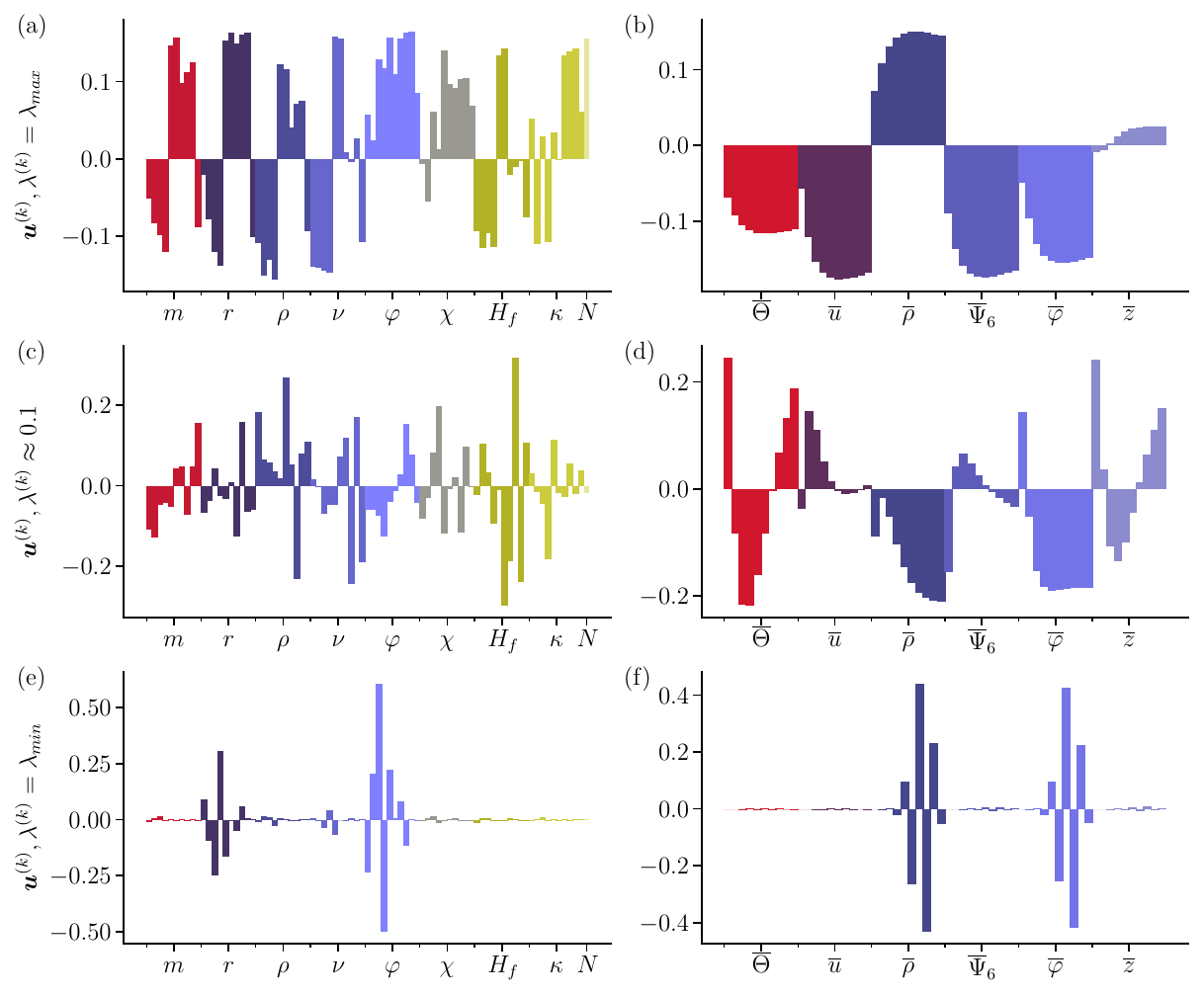}
\caption{Eigenvectors $\bm u^{(k)}$ as a function of the feature index for selected
eigenvalues of the feature correlation matrix. We show the eigenvectors
corresponding to $\lambda^{(1)}=\lambda_{\rm max}$,
$\lambda^{(k)}\approx 0.1$, and $\lambda^{(d)}=\lambda_{\rm min}$, for the
superconductivity dataset in panels~(a,c,e) and the glassy-dynamics dataset
in panels~(b,d,f).}
\label{fig:eigenvectors}
\end{figure}

We now discuss how Ridge regularization suppresses weight oscillations.
Using the eigenvalue decomposition, the Ridge
weight vector in Eq.~\eqref{eq:Ridge_solution} can be written as
\begin{equation}
    \hat{\bm w}_{\rm Ridge}
    = \sum_{k=1}^d
    \frac{1}{\lambda^{(k)}+\alpha}\,
    \bm u^{(k)} {{\bm u}^{(k)}}^T \Rb{X}{Y}.
    \label{eq:Ridge_eigenvalue_decomposition}
\end{equation}
Compared with the OLS expression, Eq.~\eqref{eq:OLS_eigenvalue_decomposition}, Ridge regularization simply replaces the factor $1/\lambda^{(k)}$ by
$1/(\lambda^{(k)}+\alpha)$. Therefore, a finite value of $\alpha$ suppresses
the amplification of modes associated with small eigenvalues,
$\lambda^{(k)}\ll \alpha$. Since these small-eigenvalue modes are responsible
for the oscillatory behavior of the OLS weights, Ridge regularization
naturally mitigates weight oscillations.

Although the regularization factor differs from that in the VIF expression,
the underlying mechanism is essentially the same as in
Eq.~\eqref{eq:VIF_eigenvalue_decomposition}. Increasing $\alpha$ suppresses
the contributions from small-eigenvalue modes, thereby reducing both the VIF
and the oscillatory components of the weights.

\subsection{Numerical tests}

To test the above theoretical arguments, we show in
Fig.~\ref{fig:eigenvectors} the eigenvectors $\bm u^{(k)}$ associated with
the eigenvalues $\lambda^{(1)}=\lambda_{\rm max}$,
$\lambda^{(k)}\approx 0.1$, and $\lambda^{(d)}=\lambda_{\rm min}$, for the
superconductivity dataset in panels~(a,c,e) and the glassy-dynamics dataset
in panels~(b,d,f). The choice $\lambda^{(k)}\simeq 0.1$ corresponds to an
eigenvalue scale comparable to the Ridge regularization strength
$\alpha=0.1$ used above. Modes with eigenvalues below this scale are therefore
strongly regularized.

We observe that, as the eigenvalue decreases, the corresponding eigenvector
becomes increasingly oscillatory as a function of the feature index. This
confirms the theoretical picture that small-eigenvalue modes are responsible
for the oscillatory behavior of the OLS weights. Ridge regularization
suppresses these modes and therefore mitigates weight oscillations.

We then numerically examine how the oscillatory behavior of the weights is suppressed by increasing $\alpha$. To this end, we define the magnitude of weight oscillations as
\begin{equation}
    S = \frac{\sum_{f=1}^{d-1} \left|\hat w^{(f)}-\hat w^{(f+1)}\right|^2}
    {\sum_{f=1}^d \left|\hat w^{(f)}\right|^2},
\end{equation}
in analogy with an elastic energy. 
$S$ corresponds to a roughness measure, which is often introduced as an additional penalty term to suppress oscillatory regression coefficients in spectroscopy regression and functional regression~\cite{gowen2011preventing,cardot2003spline}. Here, we normalize this roughness by the squared norm of the weight vector and use the resulting dimensionless quantity not as a penalty, but as a diagnostic measure to monitor the severity of weight oscillations.

\begin{figure}[b]
  \includegraphics[width=\columnwidth]{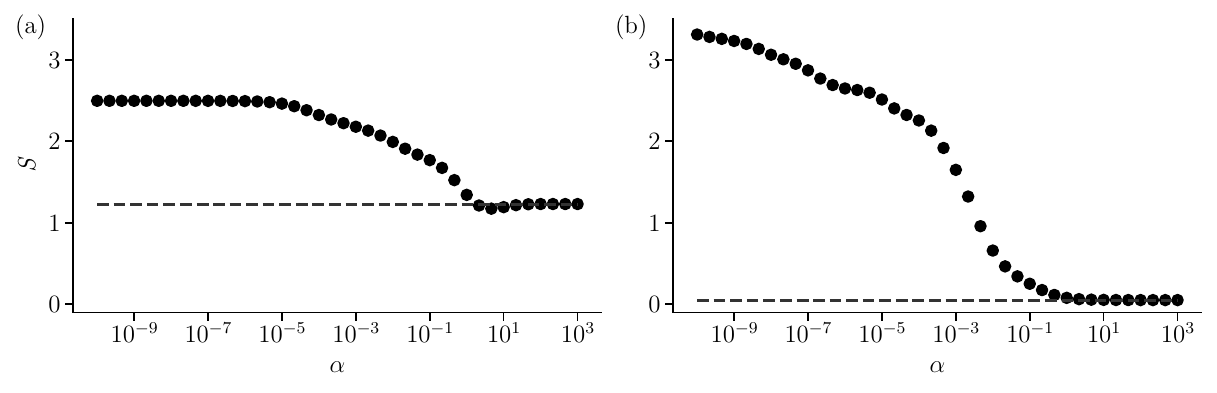}
\caption{Normalized roughness $S$ of weight oscillations as a function of the Ridge-regularization strength $\alpha$ for the superconductivity dataset in panel (a) and the glassy-dynamics dataset in panel (b). The horizontal dashed lines are the asymptotic values expected from $\hat{\bm w}_{\rm Ridge} \approx \alpha ^{-1} \Rb{X}{Y}$.}
\label{fig:weight_elastic}
\end{figure}

In Fig.~\ref{fig:weight_elastic}, we show $S$ as a function of $\alpha$ for the superconductivity dataset in panel (a) and the glassy-dynamics dataset in panel (b). We confirm that $S$ decreases steadily as $\alpha$ is increased, demonstrating that Ridge regularization suppresses oscillatory weight patterns, but only for large enough $\alpha$ values. In particular, in the glassy-dynamics dataset, the oscillations appear to decrease markedly only when $\alpha \gtrsim 10^{-3}$.

\section{Discussion and conclusions}
\label{sec:conclusion}

Because of its simplicity, linear regression remains an attractive data-driven approach in physics research, where one wishes to extract physical mechanisms from a predictive model.
However, contrary to naive expectations, even linear models can be hard to interpret~\cite{sharma2026interpretability}.
In this paper, we studied two shortcomings of linear regression that limit its interpretability: dataset-to-dataset fluctuations of the weights and oscillations of the weights across physically similar features.
These issues manifest themselves in datasets containing strongly correlated features, i.e., in the presence of multicollinearity -- a common trait of several real-life datasets.

Weight fluctuations and weight oscillations in linear regression may appear, at first sight, to be closely related phenomena.
We provided a systematic theoretical analysis that disentangles weight fluctuations from weight oscillations, while showing that both can be commonly induced by multicollinearity.
We clarified the underlying mechanisms using the eigenspace decomposition of the feature correlation matrix.
This framework also explains all the key effects of Ridge regularization, including the suppression of unstable modes associated with small eigenvalues and the largely universal behavior of the maximum variance inflation factor across a diverse range of datasets.

By itself, our analysis does not provide a practical method for extracting physically interpretable features.
Ridge regularization mitigates multicollinearity and therefore suppresses both fluctuations and oscillations, but the resulting weight pattern can depend on the magnitude of the regularization parameter $\alpha$ even when overfitting is absent and the sample size is large~\cite{montgomery2021introduction, sharma2026interpretability}.
This indicates that there is no fundamental reason to choose a unique value of $\alpha$ solely for the purpose of interpreting the weight pattern.
One may also consider principal component analysis (PCA), which removes modes associated with smaller eigenvalues by transforming the original feature vector into a smaller set of principal components.
However, because principal components are linear combinations of the original features, their physical interpretation in terms of the original variables can be less clear.

The logical conclusion is that feature selection plays a key role in data-driven research contexts where one aims to reduce model complexity 
while preserving predictive performance~\cite{boattini2021averaging, Alkemade_Boattini_Filion_Smallenburg_2022}.
Various approaches may be considered, including Lasso regression~\cite{james2013introduction}, minimum-redundancy-maximum-relevance selection~\cite{peng2005feature}, information-imbalance-based methods~\cite{glielmo2022ranking, sharma2024selecting}, and the removal of features with very large variance inflation factors.
This direction is in the same spirit as physical understanding itself: complex phenomena are often understood by compressing information into a small number of relevant variables or parameters~\cite{kivelson2018understanding}.
We leave a systematic study of which feature-selection strategies are most effective for improving interpretability to future work.

\begin{acknowledgments}
MO thanks the support by MIAI@Grenoble Alpes and the Agence Nationale de la Recherche under France 2030 with the reference ANR-23-IACL-0006).
\end{acknowledgments}

\section*{DATA AVAILABILITY}
The data and workflow necessary to reproduce the findings of this study will be available in the Zenodo data repository~\cite{zenodo} upon publication.

\appendix

\section{Diagonal elements of the inverse of the correlation matrix}
\label{sec:derivation_VIF}

We derive the relation connecting the diagonal elements of the inverse correlation matrix with the coefficient of determination obtained when a given feature is regressed on the remaining features~\cite{taboga2017lectures}.

Since the numbering of features is arbitrary, we consider $f=1$ without loss of generality. We start with
\begin{equation}
 \left( \mathrm{C}^{-1} \right)_{1, 1} 
    =  N_\mathcal{S} \left( (\mathrm{X}^T \mathrm{X})^{-1} \right)_{1,1} .
    \label{eq:variance_11}
\end{equation}
We then decompose the matrix $\mathrm{X}$ into the first feature ${\bm X}^{(1)}$ and the remaining features, namely, $\mathrm{X}=[{\bm X}^{(1)}, \mathrm{X}_{\rm R}]$,
where $\mathrm{X}_{\rm R}=[{\bm X}^{(2)},{\bm X}^{(3)},\ldots,{\bm X}^{(d)}]$ is an $N_\mathcal{S} \times (d-1)$ matrix.  
With this decomposition, the matrix $\mathrm{X}^T \mathrm{X}$ can be written as a block matrix,
\begin{equation}
    \mathrm{X}^T \mathrm{X} = 
\begin{bmatrix}
{{\bm X}^{(1)}}^T {\bm X}^{(1)} & {{\bm X}^{(1)}}^T \mathrm{X}_{\rm R} \\
\mathrm{X}_{\rm R}^T{\bm X}^{(1)}  & \mathrm{X}_{\rm R}^T\mathrm{X}_{\rm R}
\end{bmatrix}.
\end{equation}

The inverse of a block matrix can be computed using the Schur complement. In particular, the $(1,1)$ element is given by
\begin{equation}
    \left( (\mathrm{X}^T \mathrm{X})^{-1} \right)_{1,1} = S^{-1},
    \label{eq:XX_S}
\end{equation}
where $S$ is the Schur complement defined by
\begin{align}
    S &= {{\bm X}^{(1)}}^T \mathrm{M} {\bm X}^{(1)}, \\
    \mathrm{M} &= \mathrm{I} - \mathrm{X}_{\rm R} \left(\mathrm{X}_{\rm R}^T \mathrm{X}_{\rm R}\right)^{-1}\mathrm{X}_{\rm R}^T .
    \label{eq:def_M}
\end{align}
Since $\mathrm{M}$ is symmetric and idempotent, i.e., $\mathrm{M}^T=\mathrm{M}$ and $\mathrm{M}^2=\mathrm{M}$, we have
\[
S=(\mathrm{M}{\bm X}^{(1)})^T \mathrm{M}{\bm X}^{(1)}.
\]

We next consider the OLS regression of ${\bm X}^{(1)}$ on the remaining features $\mathrm{X}_{\rm R}=[{\bm X}^{(2)}, {\bm X}^{(3)}, \ldots, {\bm X}^{(d)}]$. The fitted values $\hat {\bm X}^{(1)}$ and the corresponding estimated weight vector $\hat {\bm v}_{\rm OLS}$ are given by
\begin{align}
    \hat {\bm X}^{(1)} &= \mathrm{X}_{\rm R} \hat {\bm v}_{\rm OLS}, \label{eq:hat_X1} \\
    \hat {\bm v}_{\rm OLS} &= \left(\mathrm{X}_{\rm R}^T\mathrm{X}_{\rm R}\right)^{-1}\mathrm{X}_{\rm R}^T {\bm X}^{(1)}. \label{eq:v_OLS}
\end{align}
Using Eqs.~(\ref{eq:def_M})–(\ref{eq:v_OLS}), we obtain
\[
\mathrm{M} {\bm X}^{(1)} = {\bm X}^{(1)} - \hat {\bm X}^{(1)},
\]
and therefore
\begin{equation}
S=||{\bm X}^{(1)} - \hat {\bm X}^{(1)}||^2 .
\label{eq:S_final}
\end{equation}

Since the features are normalized, the coefficient of determination in
Eq.~\eqref{eq:coefficient_determination} reduces to
\begin{equation}
    R^2[X^{(1)}, \hat X^{(1)}] 
    = 1 - \frac{1}{N_\mathcal{S}} ||{\bm X}^{(1)} - \hat {\bm X}^{(1)}||^2 .
    \label{eq:R2_appendix}
\end{equation}

Finally, combining the above equations, we conclude
\begin{equation}    \left( \mathrm{C}^{-1} \right)_{1, 1} 
    = \frac{N_\mathcal{S}}{||{\bm X}^{(1)} - \hat {\bm X}^{(1)}||^2}= \frac{1}{1-R^2[X^{(1)}, \hat X^{(1)}]} .
\end{equation}

\section{Bayesian estimation}
\label{app:Bayesian}

We discuss fluctuations (or uncertainty) of the weight ${\bm w}$ in view of Bayesian estimation. 
In the traditional frequentist view, probability is defined based on frequency of occurrence. In contrast, the Bayesian approach interprets probability as a degree of belief, which allows uncertainty to be quantified in a more flexible way.
For example, a strength of the Bayesian approach is that prior knowledge such as physical constraints can be incorporated flexibly through prior distributions.
In the frequentist framework of estimation, fluctuations originate from the randomness of the data, while the weight values are treated as fixed but unknown quantities to be estimated. The variance of the estimated weights discussed in the main text (such as the variance inflation factor) arises from dataset-to-dataset fluctuations of the point estimates. On the other hand, Bayesian estimation treats the weights as random variables and evaluates their uncertainty conditional on the observed dataset. 

\subsection{Maximum a posteriori (MAP) estimation}

The central equation of Bayesian estimation is Bayes' theorem:  
\begin{equation}
    P({\bm w}|\mathrm{X}, {\bm Y}) = 
    \frac{P({\bm Y}|{\bm w}, \mathrm{X})\,P({\bm w}|\mathrm{X})}
         {P({\bm Y}|\mathrm{X})} ,
\label{eq:bayes_formula}
\end{equation}
where $P({\bm w}|\mathrm{X}, {\bm Y})$ is the posterior distribution of ${\bm w}$ given the data $(\mathrm{X},{\bm Y})$.
In Bayesian estimation, a common point estimate for the weights is obtained by
maximum a posteriori (MAP) estimation~\cite{bishop2006pattern}, 
\begin{equation}
    \hat {\bm w}_{\rm MAP} = \arg\max_{\bm w} \; P({\bm w}|\mathrm{X}, {\bm Y}) .
\end{equation}
For MAP estimation, the denominator in Eq.~\eqref{eq:bayes_formula}, which
corresponds to the normalization factor, $P({\bm Y}|\mathrm{X}) =
    \int d{\bm w} \, P({\bm Y}|{\bm w}, \mathrm{X}) \, P({\bm w}|\mathrm{X})$
does not affect the optimization.
Besides, we define the prior distribution as $P({\bm w}) = P({\bm w} \,|\, \mathrm{X})$,
which, in the standard Bayesian setting, does not actually depend on 
$\mathrm{X}$. This reflects the assumption that $P({\bm w})$ represents our
degree of belief about ${\bm w}$ prior to observing the data.
Therefore, one only needs to consider
\begin{equation}
    P({\bm w}|\mathrm{X}, {\bm Y}) \propto
    P({\bm Y}|{\bm w}, \mathrm{X}) \, P({\bm w}) ,
    \label{eq:posterior}
\end{equation}
where $P({\bm Y}|{\bm w}, \mathrm{X})$ is the likelihood.
The main task then reduces to devising the likelihood and the prior distribution.

\subsection{Likelihood}

We assume that $\NS$ data points are sampled independently and identically (i.i.d.), so that $P({\bm Y}|{\bm w}, \mathrm{X})
    = \prod_{i=1}^{\NS} P(Y_i|{\bm w}, {\bm X}_i)$, 
where $P(Y_i|{\bm w}, {\bm X}_i)$ is given by a Gaussian distribution
with mean ${\bm w}^T {\bm X}_i$ and variance $\sigma^2$, i.e., $P(Y_i|{\bm w}, {\bm X}_i) = \mathcal{N}(Y_i \,|\, {\bm w}^T{\bm X}_i, \sigma^2)$.
It then follows that the likelihood is
\begin{equation}
    P({\bm Y}|{\bm w}, \mathrm{X})
    \propto \exp \left[ -\frac{1}{2\sigma^2}
        \, \| {\bm Y} - \mathrm{X}{\bm w} \|^2 \right] .
\end{equation}
Here the normalization factor of the Gaussian distribution is omitted,
since it does not play a role in our discussion.
The parameter $\sigma^2$ can be estimated either by the standard maximum likelihood method or within a fully Bayesian framework.

\subsection{Prior distributions}

One of the merits of Bayesian estimation is that prior information about
the weights can be incorporated into the framework. As special cases, we
will see that assuming a uniform prior distribution for ${\bm w}$ leads
to ordinary least squares (OLS) regression, while assuming a Gaussian
prior distribution for ${\bm w}$ corresponds to Ridge regression.

\vspace{0.5cm}

{\it Uniform prior:}
We consider a uniform distribution (noninformative prior)
$P({\bm w})={\rm const.}$ for ${\bm w}$. 
In this case, the posterior distribution in Eq.~\eqref{eq:posterior} becomes
\begin{equation}
    P({\bm w}|\mathrm{X}, {\bm Y})
    \propto \exp \left[ -\frac{1}{2\sigma^2}
        \, \| {\bm Y} - \mathrm{X}{\bm w} \|^2 \right] .
\end{equation}
Maximizing the (log) posterior $P({\bm w}|\mathrm{X}, {\bm Y})$ with respect
to ${\bm w}$ is therefore equivalent to minimizing the mean squared error
(MSE), $\frac{1}{2} \| {\bm Y} - \mathrm{X}{\bm w} \|^2$. Hence, the MAP estimate
reduces to ordinary least squares (OLS) regression.

\vspace{0.5cm}
{\it Gaussian prior:}
We next consider a Gaussian prior, $ P({\bm w}) \propto \exp \left[ -\frac{\NS \alpha}{2\sigma^2} {\bm w}^T {\bm w} \right]$,
which expresses the idea that the components of ${\bm w}$ are not expected
to be very large but are distributed around zero, with the variance
controlled by the hyperparameter $\alpha$. In this case, the posterior
distribution becomes
\begin{equation}
    P({\bm w}|\mathrm{X}, {\bm Y})
    \propto \exp \left[ -\frac{1}{2\sigma^2} 
        \, \| {\bm Y} - \mathrm{X}{\bm w} \|^2 
        - \frac{\NS \alpha}{2\sigma^2} {\bm w}^T {\bm w} \right] .
        \label{eq:posterior_Ridge}
\end{equation}
Maximizing $P({\bm w}|\mathrm{X}, {\bm Y})$ is therefore equivalent to
minimizing
\begin{equation}
    \frac{1}{2\NS} \| {\bm Y} - \mathrm{X}{\bm w} \|^2 
    + \frac{\alpha}{2} {\bm w}^T {\bm w} ,
\end{equation}
which shows that MAP estimation with a Gaussian prior corresponds to
Ridge regression.

\subsection{Uncertainty analysis}

The Bayesian estimation framework directly provides uncertainty
quantification of the weight estimation through the posterior
distribution. We analyze the fluctuations of the weight vector ${\bm w}$
around the MAP estimator (corresponding to OLS or Ridge in our case),
which allows us to reveal how multicollinearity, encoded in the
singularity of the correlation matrix $\mathrm{C}$, affects the estimation from
the Bayesian perspective.

Here we provide an analysis of fluctuations around the Ridge estimator, $\hat{\bm w}_{\rm Ridge} = (\mathrm{C} + \alpha \mathrm{I})^{-1} \Rb{X}{Y}$,
where $\alpha$ is the regularization parameter. This includes the OLS case
as a special limit $\alpha \to 0$.

Using Eq.~\eqref{eq:Loss_expansion}, the posterior distribution in
Eq.~\eqref{eq:posterior_Ridge} can be rewritten as
\begin{eqnarray}
    P({\bm w}|\mathrm{X}, {\bm Y})
    &\propto& \exp \left[ -\frac{\NS}{2\sigma^2} 
        ({\bm w}-\hat {\bm w}_{\rm Ridge})^T (\mathrm{C}+\alpha \mathrm{I})
        ({\bm w}-\hat {\bm w}_{\rm Ridge}) \right] \nonumber \\
    &=& \exp \left[ -\frac{1}{2} 
        ({\bm w}-\hat {\bm w}_{\rm Ridge})^T \Sigma_{\rm Ridge}^{-1}
        ({\bm w}-\hat {\bm w}_{\rm Ridge}) \right] ,
\end{eqnarray}
where $\Sigma_{\rm Ridge}$ is the posterior covariance matrix around the
Ridge estimator, given by
\begin{equation}
    \Sigma_{\rm Ridge} = \frac{\sigma^2}{\NS} (\mathrm{C} + \alpha \mathrm{I})^{-1} .
\end{equation}
In the case of the OLS estimator ($\alpha \to 0$), the covariance $\Sigma_{\rm OLS}=\frac{\sigma^2}{\NS} \mathrm{C}^{-1}$ has the singularity due to the correlation matrix. Yet, as expected, $\alpha > 0$ regularizes the covariance in the Ridge case.

\bibliography{references.bib}

@book{bishop2006pattern,
author = {Bishop, Christopher},
title = {Pattern Recognition and Machine Learning},
year = {2006},
publisher = {Springer},
isbn = {978-0-387-31073-2}
}

@book{berthier2011dynamical,
  title={Dynamical heterogeneities in glasses, colloids, and granular media},
  author={Berthier, Ludovic and Biroli, Giulio and Bouchaud, Jean-Philippe and Cipelletti, Luca and van Saarloos, Wim},
  year={2011},
  doi ={10.1093/acprof:oso/9780199691470.001.0001},
  publisher={OUP Oxford}
}

@article{glielmo2022ranking,
  title={Ranking the information content of distance measures},
  author={Glielmo, Aldo and Zeni, Claudio and Cheng, Bingqing and Cs{{\'a}}nyi, G{{\'a}}bor and Laio, Alessandro},
  journal={PNAS nexus},
  volume={1},
  pages={pgac039},
  year={2022},
  doi = {10.1093/pnasnexus/pgac039},
  publisher={Oxford University Press}
}

@article{karmakar2014growing,
  title={Growing length scales and their relation to timescales in glass-forming liquids},
  author={Karmakar, Smarajit and Dasgupta, Chandan and Sastry, Srikanth},
  journal={Annu. Rev. Condens. Matter Phys.},
  volume={5},
  pages={255},
  year={2014},
  doi = {10.1146/annurev-conmatphys-031113-133848},
  publisher={Annual Reviews}
}

@article{boattini2021averaging,
  title={Averaging local structure to predict the dynamic propensity in supercooled liquids},
  author={Boattini, Emanuele and Smallenburg, Frank and Filion, Laura},
  journal={Phys. Rev. Lett.},
  volume={127},
  pages={088007},
  year={2021},
  doi = {10.1103/PhysRevLett.127.088007},
  publisher={APS}
}

@article{sharma2024selecting,
  title={Selecting relevant structural features for glassy dynamics by information imbalance},
  author={Sharma, Anand and Liu, Chen and Ozawa, Misaki},
  journal={J. Chem. Phys.},
  volume={161},
  pages={184506},
  year={2024},
  doi = {10.1063/5.0235084},
  publisher={AIP Publishing}
}

@book{taboga2017lectures,
  title={Lectures on probability theory and mathematical statistics},
  author={Taboga, Marco},
  publisher={CreateSpace Independent Publishing Platform},
  year={2017}
}

@article{garcia2015collinearity,
  title={Collinearity: revisiting the variance inflation factor in ridge regression},
  author={Garc{\'\i}a, CB and Garc{\'\i}a, J and L{{\'o}}pez Mart{\'\i}n, MM and Salmer{{\'o}}n, R},
  journal={Journal of Applied Statistics},
  volume={42},
  pages={648},
  year={2015},
  doi = {10.1080/02664763.2014.980789},
  publisher={Taylor \& Francis}
}

@article{marquardt1970generalized,
  title={Generalized inverses, ridge regression, biased linear estimation, and nonlinear estimation},
  author={Marquardt, Donald W},
  journal={Technometrics},
  volume={12},
  pages={591},
  year={1970},
  doi = {10.1080/00401706.1970.10488699},
  publisher={Taylor \& Francis}
}

@article{mcdonald2009ridge,
  title={Ridge regression},
  author={McDonald, Gary C},
  journal={Wiley Interdisciplinary Reviews: Computational Statistics},
  volume={1},
  pages={93},
  year={2009},
  doi = {10.1002/wics.14},
  publisher={Wiley Online Library}
}

@book{montgomery2021introduction,
  title={Introduction to linear regression analysis},
  author={Montgomery, Douglas C and Peck, Elizabeth A and Vining, G Geoffrey},
  year={2021},
  publisher={John Wiley \& Sons},
  ISBN = {978-1-119-57872-7}
}

@article{jung2025roadmap,
  title={Roadmap on machine learning glassy dynamics},
  author={Jung, Gerhard and Alkemade, Rinske M and Bapst, Victor and Coslovich, Daniele and Filion, Laura and Landes, Fran{\c{c}}ois P and Liu, Andrea J and Pezzicoli, Francesco Saverio and Shiba, Hayato and Volpe, Giovanni and others},
  journal={Nat. Rev. Phys.},
  volume={7},
  pages={91},
  year={2025},
  doi = {10.1038/s42254-024-00791-4},
  publisher={Nature Publishing Group UK London}
}

@inproceedings{teney2022predicting,
  title={Predicting is not understanding: Recognizing and addressing underspecification in machine learning},
  author={Teney, Damien and Peyrard, Maxime and Abbasnejad, Ehsan},
  booktitle={European Conference on Computer Vision},
  pages={458},
  year={2022},
  doi = {10.1007/978-3-031-20050-2_27},
  organization={Springer}
}

@article{rudin2019stop,
  title={Stop explaining black box machine learning models for high stakes decisions and use interpretable models instead},
  author={Rudin, Cynthia},
  journal={Nat. Mach. Intell.},
  volume={1},
  pages={206--215},
  year={2019},
  doi={10.1038/s42256-019-0048-x},
  publisher={Nature Publishing Group UK London}
}

@article{peng2005feature,
  title={Feature selection based on mutual information criteria of max-dependency, max-relevance, and min-redundancy},
  author={Peng, Hanchuan and Long, Fuhui and Ding, Chris},
  journal={IEEE Transactions on pattern analysis and machine intelligence},
  volume={27},
  pages={1226},
  year={2005},
  doi={10.1109/TPAMI.2005.159},
  publisher={IEEE}
}

@article{kivelson2018understanding,
  title={Understanding complexity},
  author={Kivelson, Sophia and Kivelson, Steven},
  journal={Nat. Phys.},
  volume={14},
  pages={426},
  year={2018},
  doi = {10.1038/s41567-018-0136-6},
  publisher={Nature Publishing Group UK London}
}

@book{james2013introduction,
  title={An introduction to statistical learning: with applications in R},
  author={James, Gareth and Witten, Daniela and Hastie, Trevor and Tibshirani, Robert and others},
  volume={103},
  doi = {10.1007/978-1-4614-7138-7},
  year={2013},
  publisher={Springer}
}

@article{Wetzel_Ha_Iten_Klopotek_Liu_2025,
  title =	 {Interpretable Machine Learning in Physics: A Review},
  DOI =		 {10.48550/arXiv.2503.23616},
  journal =	 {arXiv preprint arXiv:2503.23616},
  author =	 {Wetzel, Sebastian Johann and Ha, Seungwoong and Iten, Raban
                  and Klopotek, Miriam and Liu, Ziming},
  year =	 2025,
}

@article{brunton2016discovering,
  title={Discovering governing equations from data by sparse identification of nonlinear dynamical systems},
  author={Brunton, Steven L and Proctor, Joshua L and Kutz, J Nathan},
  journal={Proceedings of the national academy of sciences},
  doi = {10.1073/pnas.1517384113},
  volume={113},
  number={15},
  pages={3932--3937},
  year={2016},
  publisher={National Academy of Sciences}
}

@article{rudy2017data,
  title={Data-driven discovery of partial differential equations},
  author={Rudy, Samuel H and Brunton, Steven L and Proctor, Joshua L and Kutz, J Nathan},
  journal={Science advances},
  doi = {10.1126/sciadv.1602614},
  volume={3},
  number={4},
  pages={e1602614},
  year={2017},
  publisher={American Association for the Advancement of Science}
}

@article{kac1953eigen,
  title={On the eigen-values of certain Hermitian forms},
  author={Kac, Marek and Murdock, WL and Szeg{\"o}, Gabor},
  journal={Journal of Rational Mechanics and Analysis},
  volume={2},
  pages={767--800},
  year={1953},
  publisher={JSTOR}, 
  ISSN = {19435282, 19435290},
  URL = {http://www.jstor.org/stable/24900353}
}

@article{bogoya2016eigenvectors,
  title={Eigenvectors of Hermitian Toeplitz matrices with smooth simple-loop symbols},
  author={Bogoya, Johan M and B{\"o}ttcher, A and Grudsky, Sergei M and Maximenko, Egor A},
  journal={Linear Algebra and its Applications},
  doi = {10.1016/j.laa.2015.12.017},
  volume={493},
  pages={606--637},
  year={2016},
  publisher={Elsevier}
}

@article{trench1985eigenvalue,
  title={On the eigenvalue problem for Toeplitz band matrices},
  author={Trench, William F},
  journal={Linear Algebra and its Applications},
  doi = {10.1016/0024-3795(85)90277-0},
  volume={64},
  pages={199--214},
  year={1985},
  publisher={Elsevier}
}

@article{gowen2011preventing,
  title={Preventing over-fitting in PLS calibration models of near-infrared (NIR) spectroscopy data using regression coefficients},
  author={Gowen, Aoife A and Downey, G and Esquerre, C and O'donnell, CP},
  journal={Journal of Chemometrics},
  doi = {10.1002/cem.1349},
  volume={25},
  number={7},
  pages={375--381},
  year={2011},
  publisher={Wiley Online Library}
}

@article{cardot2003spline,
 ISSN = {10170405, 19968507},
 URL = {http://www.jstor.org/stable/24307112},
 journal = {Statistica Sinica},
 number = {3},
 pages = {571--591},
 publisher = {Institute of Statistical Science, Academia Sinica},
 title = {SPLINE ESTIMATORS FOR THE FUNCTIONAL LINEAR MODEL},
 author={Cardot, Herv{\'e} and Ferraty, Fr{\'e}d{\'e}ric and Sarda, Pascal},
 urldate = {2026-06-06},
 volume = {13},
 year = {2003}
}

@article{carleo2019machine,
  title={Machine learning and the physical sciences},
  author={Carleo, Giuseppe and Cirac, Ignacio and Cranmer, Kyle and Daudet, Laurent and Schuld, Maria and Tishby, Naftali and Vogt-Maranto, Leslie and Zdeborov{\'a}, Lenka},
  journal={Reviews of Modern Physics},
  doi = {10.1103/RevModPhys.91.045002},
  volume={91},
  number={4},
  pages={045002},
  year={2019},
  publisher={APS}
}

@article{brown2009critical,
  title={Critical factors limiting the interpretation of regression vectors in multivariate calibration},
  author={Brown, Christopher D and Green, Robert L},
  journal={TrAC Trends in Analytical Chemistry},
  doi = {10.1016/j.trac.2009.02.003},
  volume={28},
  number={4},
  pages={506--514},
  year={2009},
  publisher={Elsevier}
}

@article{hamidieh2018data,
  title={A data-driven statistical model for predicting the critical temperature of a superconductor},
  author={Hamidieh, Kam},
  journal={Computational Materials Science},
  doi = {10.1016/j.commatsci.2018.07.052},
  volume={154},
  pages={346--354},
  year={2018},
  publisher={Elsevier}
}

@article{lipton2018mythos,
author = {Lipton, Zachary C.},
title = {The mythos of model interpretability},
year = {2018},
issue_date = {October 2018},
publisher = {Association for Computing Machinery},
address = {New York, NY, USA},
volume = {61},
number = {10},
issn = {0001-0782},
url = {https://doi.org/10.1145/3233231},
doi = {10.1145/3233231},
journal = {Commun. ACM},
month = sep,
pages = {36–43},
numpages = {8}
}

@article{murdoch2019definitions,
  title={Definitions, methods, and applications in interpretable machine learning},
  author={Murdoch, W James and Singh, Chandan and Kumbier, Karl and Abbasi-Asl, Reza and Yu, Bin},
  journal={Proceedings of the National Academy of Sciences},
  volume={116},
  number={44},
  pages={22071--22080},
  year={2019},
  doi = {10.1073/pnas.1900654116},
  publisher={National Academy of Sciences}
}

@article{Rowan_Doostan_2025,
  title =	 {On the definition and importance of interpretability in
                  scientific machine learning},
  DOI =		 {10.48550/arXiv.2505.13510},
  journal =	 {arXiv:2505.13510},
  author =	 {Rowan, Conor and Doostan, Alireza},
  year =	 2025,
}

@article{arrieta2020explainable,
  title={Explainable Artificial Intelligence (XAI): Concepts, taxonomies, opportunities and challenges toward responsible AI},
  author={Arrieta, Alejandro Barredo and D{\'\i}az-Rodr{\'\i}guez, Natalia and Del Ser, Javier and Bennetot, Adrien and Tabik, Siham and Barbado, Alberto and Garc{\'\i}a, Salvador and Gil-L{\'o}pez, Sergio and Molina, Daniel and Benjamins, Richard and others},
  journal={Information fusion},
  doi = {10.1016/j.inffus.2019.12.012},
  volume={58},
  pages={82--115},
  year={2020},
  publisher={Elsevier}
}

@article{gunning2019xai,
  title={XAI—Explainable artificial intelligence},
  author={Gunning, David and Stefik, Mark and Choi, Jaesik and Miller, Timothy and Stumpf, Simone and Yang, Guang-Zhong},
  journal={Sci. Robot.},
  doi = {10.1126/scirobotics.aay7120},
  volume={4},
  number={37},
  pages={eaay7120},
  year={2019},
  publisher={American Association for the Advancement of Science}
}

@article{fisher2019all,
  title={All models are wrong, but many are useful: Learning a variable's importance by studying an entire class of prediction models simultaneously},
  author={Fisher, Aaron and Rudin, Cynthia and Dominici, Francesca},
  journal={Journal of Machine Learning Research},
  volume={20},
  number={177},
  pages={1--81},
  year={2019},
  url = {http://jmlr.org/papers/v20/18-760.html}

}

@article{takahama2015model,
  title={Model selection for partial least squares calibration and implications for analysis of atmospheric organic aerosol samples with mid-infrared spectroscopy},
  author={Takahama, Satoshi and Dillner, Ann M},
  journal={Journal of Chemometrics},
  doi = {10.1002/cem.2761},
  volume={29},
  number={12},
  pages={659--668},
  year={2015},
  publisher={Wiley Online Library}
}

@article{sharma2026interpretability,
  title={Interpretability of linear regression models of glassy dynamics},
  author={Sharma, Anand and Liu, Chen and Ozawa, Misaki and Coslovich, Daniele},
  journal={Physical Review Materials},
  doi = {10.1103/q6pd-7trs},
  volume={10},
  number={3},
  pages={035602},
  year={2026},
  publisher={APS}
}

@article{austin2015number,
  title={The number of subjects per variable required in linear regression analyses},
  author={Austin, Peter C and Steyerberg, Ewout W},
  journal={Journal of clinical epidemiology},
  volume={68},
  number={6},
  doi = {10.1016/j.jclinepi.2014.12.014},
  pages={627--636},
  year={2015},
  publisher={Elsevier}
}

@article{green1991many,
  title={How many subjects does it take to do a regression analysis},
  author={Green, Samuel B},
  journal={Multivariate behavioral research},
  doi = {10.1207/s15327906mbr2603\_7},
  volume={26},
  number={3},
  pages={499--510},
  year={1991},
  publisher={Taylor \& Francis}
}

@article{lee2016prediction,
  title={Prediction model of band gap for inorganic compounds by combination of density functional theory calculations and machine learning techniques},
  author={Lee, Joohwi and Seko, Atsuto and Shitara, Kazuki and Nakayama, Keita and Tanaka, Isao},
  journal={Physical Review B},
  doi = {10.1103/PhysRevB.93.115104},
  volume={93},
  number={11},
  pages={115104},
  year={2016},
  publisher={APS}
}

@article{seko2014machine,
  title={Machine learning with systematic density-functional theory calculations: Application to melting temperatures of single-and binary-component solids},
  author={Seko, Atsuto and Maekawa, Tomoya and Tsuda, Koji and Tanaka, Isao},
  journal={Physical Review B},
  doi = {10.1103/PhysRevB.89.054303},
  volume={89},
  number={5},
  pages={054303},
  year={2014},
  publisher={APS}
}

@article{fransson2020efficient,
  title={Efficient construction of linear models in materials modeling and applications to force constant expansions},
  author={Fransson, Erik and Eriksson, Fredrik and Erhart, Paul},
  journal={npj Computational Materials},
  doi = {10.1038/s41524-020-00404-5},
  volume={6},
  number={1},
  pages={135},
  year={2020},
  publisher={Nature Publishing Group UK London}
}

@article{marchand2023multiscale,
  title={Multiscale data-driven energy estimation and generation},
  author={Marchand, Tanguy and Ozawa, Misaki and Biroli, Giulio and Mallat, St{\'e}phane},
  journal={Physical Review X},
  doi = {10.1103/PhysRevX.13.041038},
  volume={13},
  number={4},
  pages={041038},
  year={2023},
  publisher={APS}
}

@book{Thom_2016,
  title =	 {{To Predict is Not to Explain: Conversations on Mathematics,
                  Science, Catastrophe Theory, Semiophysics, Natural Philosophy
                  and Morphogenesis}},
  ISBN =	 {978-0-9939269-2-1},
  publisher =	 {Thombooks Press},
  author =	 {Thom, Rene},
  year =	 2016,
}

@article{james2009functional,
  title =	 {Functional linear regression that’s interpretable},
  volume =	 37,
  ISSN =	 {0090-5364, 2168-8966},
  DOI =		 {10.1214/08-AOS641},
  number =	 {5A},
  journal =	 {The Annals of Statistics},
  publisher =	 {Institute of Mathematical Statistics},
  author =	 {James, Gareth M. and Wang, Jing and Zhu, Ji},
  year =	 2009,
  month =	 oct,
  pages =	 {2083–2108},
}

@misc{zenodo,
  title =	 {The limits of interpretability in multiple linear regression},
  author =	 {Sharma, Anand and Liu, Chen and Coslovich, Daniele and Ozawa, Misaki},
  year =	 2026,
  publisher =	 {Zenodo},
  howpublished = "\url{https://doi.org/10.5281/zenodo.20648797}",
}

@book{brockwell2002introduction,
  title={Introduction to time series and forecasting},
  author={Brockwell, Peter J and Davis, Richard A},
  year={2002},
  publisher={Springer}, 
}

@misc{el_nino_122,
  key          = {UCI},
  title        = {El~Ni\~no},
  year         = {1999},
  howpublished = {UCI Machine Learning Repository},
  doi       = {10.24432/C5WG62},
  url       = {https://doi.org/10.24432/C5WG62}
}

@article{Wine_paper,
title = {Modeling wine preferences by data mining from physicochemical properties},
author={Paulo Cortez and Antonio Lu{\'i}z Cerdeira and Fernando Almeida and Telmo Matos and Jos{\'e} Reis},
journal = {Decision Support Systems},
volume = {47},
number = {4},
pages = {547-553},
year = {2009},
issn = {0167-9236},
doi = {/10.1016/j.dss.2009.05.016},
url ={https://www.sciencedirect.com/science/article/pii/S0167923609001377},
}

@misc{wine_dataset,
  key          = {UCI},
  title        = {Wine Quality},
  year         = {2009},
  howpublished = {UCI Machine Learning Repository},
  doi       = {10.24432/C56S3T},
  url       = {https://doi.org/10.24432/C56S3T}
}

@misc{infrared_dataset,
  key          = {UCI},
  title        = {Infrared Thermography Temperature},
  year         = {2023},
  howpublished = {UCI Machine Learning Repository},
  doi       = {10.13026/9ay4-2c37},
  url       = {https://doi.org/10.13026/9ay4-2c37}
}

@article{infrared_paper,
  author    = {Wang, Quanzeng and Zhou, Yangling and Ghassemi, Pejman
               and McBride, David and Casamento, Jon P. and Pfefer, T. Joshua},
  title     = {Infrared Thermography for Measuring Elevated Body Temperature:
               Clinical Accuracy, Calibration, and Evaluation},
  journal   = {Sensors},
  volume    = {22},
  number    = {1},
  pages     = {215},
  year      = {2022},
  publisher = {MDPI},
  doi       = {10.3390/s22010215},
  url       = {https://doi.org/10.3390/s22010215}
}

@misc{appliance_energy_data,
  key          = {UCI},
  title        = {Appliances Energy Prediction},
  year         = {2017},
  howpublished = {UCI Machine Learning Repository},
  doi       = {10.24432/C5VC8G},
  url       = {https://doi.org/10.24432/C5VC8G}
}

@article{appliance_energy_paper,
title = {Data driven prediction models of energy use of appliances in a low-energy house},
author = {Luis M. Candanedo and V\'eronique Feldheim and Dominique Deramaix},
journal = {Energy and Buildings},
volume = {140},
pages = {81-97},
year = {2017},
issn = {0378-7788},
doi = {10.1016/j.enbuild.2017.01.083},
url = {https://www.sciencedirect.com/science/article/pii/S0378778816308970}
}

@InProceedings{online_news_paper,
author={Fernandes, Kelwin
and Vinagre, Pedro
and Cortez, Paulo},
title={A Proactive Intelligent Decision Support System for Predicting the Popularity of Online News},
booktitle="Progress in Artificial Intelligence",
year={2015},
publisher={Springer},
address={Cham},
pages={535--546},
isbn={978-3-319-23485-4}, 
doi = {10.1007/978-3-319-23485-4_53}
}

@misc{online_news_data,
  key          = {UCI},
  title        = {Online News Popularity},
  year         = {2015},
  howpublished = {UCI Machine Learning Repository},
  doi       = {10.24432/C5NS3V},
  url       = {https://doi.org/10.24432/C5NS3V}
}

@article{Alkemade_Smallenburg_Filion_2023,
  title =	 {Improving the prediction of glassy dynamics by pinpointing the
                  local cage},
  volume =	 158,
  DOI =		 {10.1063/5.0144822},
  number =	 13,
  journal =	 {J. Chem. Phys.},
  author =	 {Alkemade, Rinske M. and Smallenburg, Frank and Filion, Laura},
  year =	 2023,
  month =	 apr,
  pages =	 134512
}

@article{Alkemade_Boattini_Filion_Smallenburg_2022,
  title =	 {Comparing machine learning techniques for predicting glassy
                  dynamics},
  volume =	 156,
  DOI =		 {10.1063/5.0088581},
  number =	 20,
  journal =	 {J. Chem. Phys.},
  author =	 {Alkemade, Rinske M. and Boattini, Emanuele and Filion, Laura
                  and Smallenburg, Frank},
  year =	 2022,
  month =	 may,
  pages =	 204503
}

\end{document}